\documentclass[iop]{emulateapj}

\usepackage[caption=false]{subfig} 

\slugcomment{Accepted by the Astronomical Journal on 2015 Dec 03}

\shorttitle{Cool Dust \& Hypergiant Mass Loss}
\shortauthors{Shenoy et al.}

\begin{document}

\title{Searching for Cool Dust in the Mid-to-Far Infrared:  the Mass Loss Histories of The Hypergiants $\mu$ Cep, VY CMa, IRC+10420, and $\rho$ Cas$^{\dag}$} 

\author{Dinesh Shenoy\altaffilmark{1},Roberta M. Humphreys\altaffilmark{1}, Terry J. Jones\altaffilmark{1}, Massimo Marengo\altaffilmark{2}, Robert D. Gehrz\altaffilmark{1}, L. Andrew Helton\altaffilmark{3}, William F. Hoffmann\altaffilmark{4}, Andrew J. Skemer\altaffilmark{4}, Philip M. Hinz\altaffilmark{4}} 

\altaffiltext{\dag}{Based on observations obtained with:  (1) the NASA/DLR Stratospheric Observatory for Infrared Astronomy (SOFIA). SOFIA is jointly operated by the Universities Space Research Association, Inc. (USRA), under NASA contract NAS2-97001, and the Deutsches SOFIA Institut (DSI) under DLR contract 50 OK 0901 to the University of Stuttgart; and (2) the MMT Observatory on Mt. Hopkins, AZ, a joint facility of the Smithsonian Institution and the University of Arizona.}
\altaffiltext{1}{Minnesota Institute for Astrophysics, School of Physics and Astronomy, University of Minnesota, 116 Church St. SE, Minneapolis, MN 55455, USA,  shenoy@astro.umn.edu }
\altaffiltext{2}{Department of Physics, Iowa State University, Ames, IA 50011, USA }
\altaffiltext{3}{USRA-SOFIA Science Center, NASA Ames Research Center, Moffett Field, CA 94035, USA }
\altaffiltext{4}{Department of Astronomy/Steward Observatory, University of Arizona, 933 N. Cherry Ave, Tucson AZ 85721, USA }

\begin{abstract}
We present mid- and far- IR imaging of four famous hypergiant stars:  the red supergiants $\mu$ Cep and VY CMa, and the warm hypergiants IRC +10420 and $\rho$ Cas.  Our 11 to 37 $\micron$ SOFIA/FORCAST imaging probes cool dust not detected in visual and near-IR imaging studies.  Adaptive optics (AO) 8 - 12 $\micron$ imaging of $\mu$ Cep and IRC +10420 with MMT/MIRAC reveals extended envelopes that are the likely sources of these stars' strong silicate emission features.  We find $\mu$ Cep's mass-loss rate to have declined by about a factor of 5 over a 13,000 history, ranging from 5 $\times$ 10$^{-6}$ down to $\sim$1 $\times$ 10$^{-6}$ $M_{\sun}$ yr$^{-1}$.  The morphology of VY CMa indicates a cooler dust component coincident with the highly asymmetric reflection nebulae seen in the visual and near-IR.  The lack of cold dust at greater distances around VY CMa indicates its mass-loss history is limited to the last $\sim$1200 years, with an average rate of 6 $\times$ 10$^{-4}$ $M_{\sun}$ yr$^{-1}$.  We find two distinct periods in the mass-loss history of IRC +10420 with a high rate of 2 $\times$ 10$^{-3}$ $M_{\sun}$ yr$^{-1}$ until approximately 2000 yr ago, followed by an order of magnitude decrease in the recent past.  We interpret this change as evidence of its evolution beyond the RSG stage.  Our new infrared photometry of $\rho$ Cas is consistent with emission from the expanding dust shell ejected in its 1946 eruption, with no evidence of newer dust formation from its more recent events. 
\end{abstract} 


\section{Introduction}
The fate of massive stars is strongly affected by their mass-loss rates and mass-loss histories.  The majority of stars above 9 $M_{\sun}$ will pass through the RSG stage, during which they shed large amounts of mass.  Depending on the duration and rate of mass-loss, the total mass shed during the RSG stage can represent a significant fraction of the initial mass of the star and influences its terminal state, whether as a supernova or in post-RSG evolution.  However, the mechanism by which RSGs lose mass remains uncertain.  For some RSGs and post-RSGs we find evidence of discrete, episodic mass-loss.    For example, high resolution optical imaging and spectroscopy of the warm OH/IR post-red supergiant (RSG) IRC +10420 and the peculiar OH/IR M-type supergiant VY CMa have yielded surprising results about their complex circumstellar environments, with evidence for asymmetric ejections and multiple high mass-loss events \citep{Smith:2001,Humphreys:2005,Humphreys:2007, Jones:2007, Tiffany:2010, Shenoy:2013}.

Since a substantial portion of the emission from dusty ejected material is thermal, observations in the infrared are important for mapping the extent of mass loss and the mass-loss history.  In this paper we report on a search over a range of wavelengths for extended warm and cold dust around four famous hypergiant stars:  the RSGs $\mu$ Cep and VY CMa and the warm hypergiants IRC +10420 and $\rho$ Cas.  Thermal emission from dust in their circumstellar environment may be evidence of recent high mass loss episodes or may be a fossil record of earlier mass loss.   We extend the study of these stars' mass-loss into the mid-IR, reporting new SOFIA/FORCAST \citep{Herter:2012} 11 - 37 $\micron$ imaging, combined with publicly available \emph{Herschel}\footnote{\footnotesize{~\emph{Herschel} is an ESA space observatory with science instruments provided by European-led Principal Investigator consortia and with important participation from NASA.  The \emph{Herschel} data used in this paper are taken from the Level 2 (flux-calibrated) images provided by the Herschel Science Center via the NASA/IPAC Infrared Science Archive (IRSA), which is operated by the Jet Propulsion Laboratory, California Institute of Technology, under contract with NASA.}} \citep{Pilbratt:2010} PACS \citep{Poglitsch:2010} images.  We also include sub-arcsecond resolution 8 - 12 $\micron$ observations of $\mu$ Cep, IRC +10420, and $\rho$ Cas made with MMT/MIRAC \citep{Hoffmann:1998, Hinz:2000}.  The observations are described in \S{2} and are summarized in Tables 1 and 2.  We discuss our results in \S{3} and summarize them in Table 3 and in the last section.

\section{Observations and Data Reduction}
\subsection{SOFIA/FORCAST:  Far-IR Imaging (11 $-$ 37 $\micron$)}
We observed $\mu$ Cep, VY CMa, IRC +10420 and $\rho$ Cas with SOFIA/FORCAST during Cycle 2 (OBS ID 02\_0031, PI: R.M. Humphreys).  FORCAST is a dual-channel mid-IR imager  covering the 5 to 40 $\micron$ range.   Each channel uses a 256 $\times$ 256 pixel blocked-impurity-band (BiB) array and provides a distortion-corrected 3$\farcm$2 $\times$ 3$\farcm$2 field of view with a scale of 0$\farcs$768 pixel$^{-1}$.  FORCAST achieves near-diffraction limited imaging, with a point-spread function (PSF) full-width-at-half-maximum (FWHM) of $\sim$ 3$\farcs$7 in the longest filters.  The images were obtained in single-beam mode to maximize throughput for detecting faint extended emission.  The observations were made in standard two-position chop-and-nod mode with the direction of the nod matching the direction of the chop.  The data were reduced by the staff of the SOFIA Science Center using the FORCAST Redux v1.0.1 pipeline.  After correction for bad pixels and droop effects, the pipeline removed sky and telescope background emission by first subtracting chopped image pairs and then subtracting nodded image pairs.  The resulting positive images are aligned and merged.  The details of the FORCAST pipeline are discussed in the Guest Investigator Handbook for FORCAST Data Products, Rev. A2.\footnote{\footnotesize{Available at http://www.sofia.usra.edu/Science/index.html under ``SOFIA Data Products'' }}

Bright point sources cause cross-talk in the horizontal direction on the FORCAST array.  To mitigate this effect, chop angles were selected so that the cross-talk pattern from one chop position did not overlap with the other chop position.  Additionally, the FORCAST pipeline applies a correction that reduces the effect, although some of the pattern remains for some targets.  The effect is most noticeable for the two brightest targets, VY CMa and IRC +10420; it is less for $\mu$ Cep and is not present in the images of $\rho$ Cas.  For VY CMa and IRC +10420 we 
avoid the array horizontal direction when examining the images for evidence of extended emission.  For $\mu$ Cep where the effect is less apparent, we have marked on the images where the faint pattern is present.  Calibrator star observations from each flight were obtained from the SOFIA Science Archive for comparison.  We display the calibrators observed on the same flight as our science targets and in the same filter with the four-position slide in either the mirror position (for the short wavelength channel) or the open position (for the long wavelength channel).  In some instances the merging of the two chop beams by the FORCAST data reduction pipeline imparted an extended shape to the calibrator star that is not present in at least one or sometimes both of the two individual pre-merge images of the calibrator star.  When this effect was apparent we have displayed one of the pre-merge images for comparison.  This blurring effect during this merging of the two chop beams does not appear to have occurred in our final merged science images, for which we compared each of the pre-merge images to the merged image as well.  In particular, the extended shapes of both VY CMa and IRC +10420 discussed below are not an effect of the merging step since the extended shape is clearly seen in the individual chop beams.

The FORCAST pipeline coadds the merged images.  To determine a 1-$\sigma$ uncertainty in our quoted fluxes, for each of our targets we computed the standard deviation of the mean of fluxes extracted from the merged images prior to coadding. For the fluxes of the bright targets $\mu$ Cep, VY CMa and IRC +10420, the fractional variation in total ADUs in each filter is negligible compared to the 6\% uncertainty in the flux calibration per the GI Handbook \S 4.1 \citep{Herter:2013}.  Therefore for those targets we adopt 6\% as the 1-$\sigma$ uncertainty.  For $\rho$ Cas the fractional variations are between 1 $-$ 6\% and are added in quadrature with the 6\% uncertainty from the flux calibration to be the 1-$\sigma$ uncertainty.  The band-passes of the selected FORCAST filters are such that only small color corrections are required.  Based on the mostly $\lambda F_{\lambda}\propto\lambda^{-3}$ spectral shapes of our targets in the relevant range, we have applied small color corrections\footnote{\footnotesize{Obtained from http://www.sofia.usra.edu/Science/\\DataProducts/FORCAST\_ColorCorrecns\_OC1.pdf}} of 1.004, 1.071, 1.004, 1.044, 1.025, and 1.025 to fluxes extracted from the F111, F197, F253, F315, F348 and F371 images respectively.  The FORCAST observations are summarized in Table 1.

\begin{deluxetable*}{ccccc}
\tablecaption{List of SOFIA FORCAST Observations}	
\tablenum{1}
\tablehead{\colhead{Target} & \colhead{Date Obs} & \colhead{Filters\tablenotemark{a}} & \colhead{Int Time} & \colhead{PSF FWHM} \\ 
\colhead{} & \colhead{(UT)} & \colhead{} & \colhead{(s)} & \colhead{(arcsec)} } 
\startdata
$\mu$ Cep &  2014 05 03 & F111, F197, F253 & 14, 12, 343 & 2$\farcs$6, 2$\farcs$6, 3$\farcs$6 \\
 &  &  F315, F348, F371 & 521, 1034, 1865 & 3$\farcs$2, 3$\farcs$5, 3$\farcs$7 \\
 &  & & \\
VY CMa & 2014 03 22 & F197, F253  & 36, 284  & 2$\farcs$7, 2$\farcs$9 \\
 &  & F315, F348, F371 & 320, 856, 392 & 3$\farcs$2, 3$\farcs$5, 3$\farcs$7 \\
 & & & \\
IRC +10420 &  2014 06 06 & F197, F253  & 18, 215 &  3$\farcs$2, 3$\farcs$2 \\
 &  &  F315, F348, F371 & 283, 592, 989 & 3$\farcs$5, 3$\farcs$7, 3$\farcs$8 \\
 & & & \\
$\rho$ Cas &  2014 03 27 & F197, F253  & 42, 237  & 3$\farcs$0, 3$\farcs$3 \\
 & &  F315, F348, F371 & 319, 848, 1821 & 3$\farcs$4, 3$\farcs$7, 4$\farcs$0 \\
\enddata
\tablenotetext{a}{~The effective wavelengths of the SOFIA FORCAST filters are: F111 = 11.1 $\micron$ ($\Delta\lambda$ = 0.95 $\micron$),  F197 = 19.7 $\micron$ ($\Delta\lambda$ = 5.5 $\micron$), F253 = 25.3 $\micron$ ($\Delta\lambda$ = 1.86 $\micron$), F315 = 31.5 $\micron$ ($\Delta\lambda$ = 5.7 $\micron$), F348 = 34.8 $\micron$ ($\Delta\lambda$ = 3.8 $\micron$), F371 = 37.1 $\micron$  ($\Delta\lambda$ = 3.3 $\micron$).  Source:  SOFIA Observer's Handbook v4.1.0 \S7.1.3.\\}
\end{deluxetable*}

\subsection{MIRAC3/4:  Adaptive Optics Mid-IR Imaging\\(8 $-$ 12 $\micron$)}
Three of our SOFIA targets were also observed at high angular resolution in the 8 $-$ 12 $\micron$ range using the MIRAC3/4 cameras on the 6.5-m MMT  \citep{Hoffmann:1998, Hinz:2000}.  $\mu$ Cep was observed on UT 2006 July 22 with MIRAC3, while $\rho$ Cas, IRC +10420 and  additional RSG targets were observed with  MIRAC4 \citep{Skemer:2008} between 2006 November 05 and 2009 October 02.  Both MIRAC configurations achieved very high Strehl-ratios (approx. 0.95), providing diffraction-limited imaging and stable PSFs \citep[e.g.,][]{Biller:2005}.  MIRAC3 and MIRAC4 employed Si:As arrays with 128 $\times$ 128 and 256 $\times$ 256 pixels respectively.  Observations were made with a standard chop-and nod sequence to remove IR background emission.  Cross-talk in the array electronics introduced faint artifacts in the horizontal and vertical directions that were not completely removed by chop-and-nod subtraction.  For the MIRAC4 images the horizontal cross-talk is mitigated during the data reduction with a code written by M. Marengo \citep{Skemer:2008}.  Extension in directions other than array vertical and horizontal are not affected by cross-talk.   The PSF-subtracted images of $\mu$ Cep presented here are reproduced from \citet{Schuster:2007}, while the images of IRC +10420 are previously unpublished.  The additional RSG targets RW Cep, RW Cyg and BD +24 3902 resemble IRC +10420 with a similar extended appearance that is azimuthally symmetric, while $\rho$ Cas and W Per are point-like with no noticeable extension compared to the PSF stars.  S Per is extended at 8.9 and 9.8 micron with an elliptical envelope oriented in a NNE-SSW direction similar to extension seen in the \emph{HST} visual images \citep{Schuster:2006}.  T Per is marginally extended at 11.9 $\micron$.     Flux calibration for the 9.8 and 11.7 $\micron$ images of $\mu$ Cep was done using observations of the PSF standard $\gamma$ Dra taken immediately afterward, with an estimated uncertainty of 10\% in the calibration.  Flux calibration of the $\rho$ Cas observations was done using the standard stars $\beta$ And, $\alpha$ Aur and $\beta$ Peg, with an estimated uncertainty of 10\% as well.  The MIRAC observations are summarized in Table 2.  

\begin{deluxetable*}{cccccc}
\tablecaption{List of MMT/MIRAC\tablenotemark{*} Observations}	
\tablenum{2}
\tablehead{\colhead{Target} & \colhead{Date Obs} & \colhead{Filter(s)\tablenotemark{a}} & \colhead{Int Time} & \colhead{PSF FWHM} & \colhead{Extension} \\ 
\colhead{} & \colhead{(UT)} & \colhead{($\micron$)} & \colhead{(s)} & \colhead{(arcsec)} & \colhead{evident?} } 
\startdata
$\mu$ Cep &  2006 07 23 & 8.8, 9.8, 11.7 & 272, 362, 483 &  0$\farcs$34, 0$\farcs$38, 0$\farcs$45 & Yes (Fig \ref{muCep_MIRAC3})\\
 &  & & \\
$\rho$ Cas &  2006 11 05 & 8.9, 9.8  & 156, 331 & 0$\farcs$34, 0$\farcs$37 & No \\
 & & & \\
S Per         &  2006 11 05 & 8.9, 9.8  & 26, 260 & 0$\farcs$34, 0$\farcs$37 & Yes \\
 & & & \\
IRC +10420 &  2008 06 16 & 8.9, 9.8, 11.9 & 289, 313, 337 &  0$\farcs$34, 0$\farcs$37, 0$\farcs$47 & Yes (Fig \ref{IRC_MIRAC4}) \\
 & & & \\
RW Cep &  2009 09 30 & 8.7, 9.8, 11.9  & 1094, 1094, 1064 & 0$\farcs$38, 0$\farcs$43, 0$\farcs$52 & Yes\tablenotemark{b} \\
 & & & \\
W Per &  2009 09 30 & 8.7, 9.8, 11.9  & 1117, 764, 265 & 0$\farcs$38, 0$\farcs$43, 0$\farcs$52 & No \\
 & & & \\
RW Cyg &  2009 10 01 & 8.7, 9.8, 11.9  & 1324, 1204, 1385 & 0$\farcs$39, 0$\farcs$42, 0$\farcs$53 & Yes\tablenotemark{b} \\
 & & & \\
BD +24 3902 &  2009 10 02 & 8.7, 9.8, 11.9  & 1440, 1440, 1140 & 0$\farcs$39, 0$\farcs$42, 0$\farcs$53 & Yes\tablenotemark{b} \\
 & & & \\
T Per             &  2009 10 02 & 8.7, 9.8, 11.9  &  202, 193, 101 & 0$\farcs$39, 0$\farcs$42, 0$\farcs$53 & Yes\tablenotemark{b} \\
\enddata
\tablenotetext{*}{~The 2006 July observations of $\mu$ Cep were made with the MIRAC3 instrument.  Observations starting 2006 November were made with the successor instrument MIRAC4.  See \S2.2 for details.}
\tablenotetext{a}{~The bandwidths of the MIRAC3 8.8, 9.8 and 11.7 $\micron$ filters are $\Delta\lambda$ = 0.88, 0.98 and 1.12 $\micron$ respectively.  The bandwidths of the MIRAC4 8.9, 9.8 and 11.9 filters are $\Delta\lambda$ = 1.22, 0.91, and 1.14 $\micron$ respectively.}
\tablenotetext{b}{~The extended shape apparent after PSF subtraction is azimuthally symmetric, resembling that of IRC +10420 in Figure \ref{IRC_MIRAC4} (bottom row).\\}
\end{deluxetable*}

\subsection{\emph{Herschel}/PACS}
We also include in our analysis the publicly available 70 \& 160 $\micron$ observations of $\mu$ Cep, VY CMa and IRC +10420 made with \emph{Herschel}/PACS as part of the \emph{Herschel} key program MESS \citep[$M$ass-loss of $E$volved $S$tar$S$,][]{Groenewegen:2011}.\footnote{\footnotesize{~The data are taken from Obs IDs 1342191945 \& 1342191946 ($\mu$ Cep, observed on 2010 Mar 10); 1342194070 \& 1342194071 (VY CMa, observed on 2010 Apr 08) and 1342196028 \& 1342196029 (IRC +10420, observed on 2010 May 09).}}  The width of the PACS bandpasses require color corrections be applied to the 70 \& 160 $\micron$ photometry taken from the images.  To compute the necessary corrections we convolve the PACS ``blue'' (70 $\micron$) and ``red'' (160 $\micron$) response functions\footnote{\footnotesize{~http://herschel.esac.esa.int/twiki/pub/Public/Pacs\\CalibrationWeb/cc\_report\_v1.pdf}} with each target's ISO LWS \citep{Clegg:1996} spectrum.  For the blue bandpass we use the LWS spectrum itself; for the red bandpass we represent the target's SED as a power law of the form $F_{\nu} \propto \nu^{\beta}$ since the LWS spectrum does not cover the full red bandpass.   At 70 and 160 $\micron$ respectively we compute and apply corrections of:  1.045 and 1.064 for $\mu$ Cep; 1.031 and 1.099 for VY CMa; 1.002 and 1.080 for IRC +10420.

\section{Results \& Discussion}
\subsection{Method for Estimating Mass-Loss Rates}
We compute mass-loss rates for the resolved targets $\mu$ Cep, VY CMa and IRC +10420 using the DUSTY  radiative transfer code \citep{Ivezic:1997} to model the stars' spectral energy distributions (SEDs) and azimuthal average intensity profiles at one or more wavelengths.  DUSTY solves the radiative transfer equation for a spherically symmetric dust distribution around a central source.  The user specifies the spectrum of the illuminating source, the optical properties and size distribution of the dust grains, the dust temperature at radius $r_{1}$ (the inner boundary of the shell), a functional form for the radial profile of the dust density $\rho(r)$ throughout the shell from inner radius $r_{1}$ to outer radius $r_{2}$, and the total optical depth at a chosen wavelength for a line of sight through the shell directly to the central source.  For a given set of these inputs, DUSTY outputs a model SED and a radial profile of the intensity of the shell at desired wavelengths.  

Assuming spherical symmetry is a simplification that cannot capture the full 3-D spatial distribution of asymmetric circumstellar ejecta, which is known to be particularly complex in the optical and near-infrared in the cases of VY CMa \citep{Smith:2001, Humphreys:2005,Humphreys:2007, Shenoy:2013} and IRC +10420 \citep{Humphreys:1997,Tiffany:2010, Shenoy:2015}.  The effect of asymmetry is less influential at lower optical depths, which is the case for $\mu$ Cep (see \S 3.2.2 below).  The advantage gained by the use of a radiative transfer code such as DUSTY is a relatively consistent treatment across our targets, allowing for comparison of the results for similar assumptions about the dust responsible for their infrared SEDs and intensity profiles.    

For a fitted model's optical depth $\tau_{\lambda}$, the grain opacity $\kappa_{\lambda}$ specifies the dust mass density $\rho(r)$ in g cm$^{-3}$ throughout the shell.  If the expansion velocity is constant throughout the shell, the mass-loss rate at a time $t$ in the past is the rate at which gas and dust pass radius $r$ $=$ $v_{exp}\cdot t$.  This may be written as $\dot{M}(t) = g_{d}\cdot4\pi r^{2} \cdot \rho(r) \cdot v_{exp}$ where $g_{d}$ is the gas-to-dust mass ratio, which we assume to be 100:1.\footnote{\footnotesize{~The gas-to-dust mass ratios assumed or derived for these targets typically range from 100:1 \citep[e.g.,][]{Knapp:1993, Verhoelst:2009} to 200:1 \citep[e.g.,][]{Mauron:2011,Cox:2012,Oudmaijer:1996}.}}  For this constant expansion velocity case, a model with density distribution $\rho(r)\propto r^{-q}$ with power law index $q$ $<$ 2  indicates a gradual decline in the mass-loss rate over the dynamical age of the shell.  

If the expansion velocity has not been constant at all times throughout the shell, we compute an average mass-loss rate.  We multiply the density distribution $\rho(r)$ by the same assumed 100:1 gas-to-dust mass ratio and integrate to compute $M$, the total mass of gas and dust in the shell.  We assume an average expansion velocity to compute the dynamical age of the shell $\Delta t$ $=$ $r_{2}$ / $v_{exp}$ where $r_{2}$ is the outer radius of the shell for the model.  The average mass loss rate is then  $\langle\dot{M}\rangle$ $=$ $M$ / $\Delta t$.   The particular choices for grain type, size distribution, shell radii and dust density distribution are discussed in the sections specific to each target below, as well as the limitations on assuming a single average expansion velocity in the cases of VY CMa and IRC +10420.  The $\langle\dot{M}\rangle$ values determined for $\mu$ Cep, VY CMa and IRC +10420 are summarized in Table 3.

\subsection{$\mu$ Cep}
\subsubsection{Mid-IR Imaging}
\begin{figure}
\begin{center}
\includegraphics[scale=0.25]{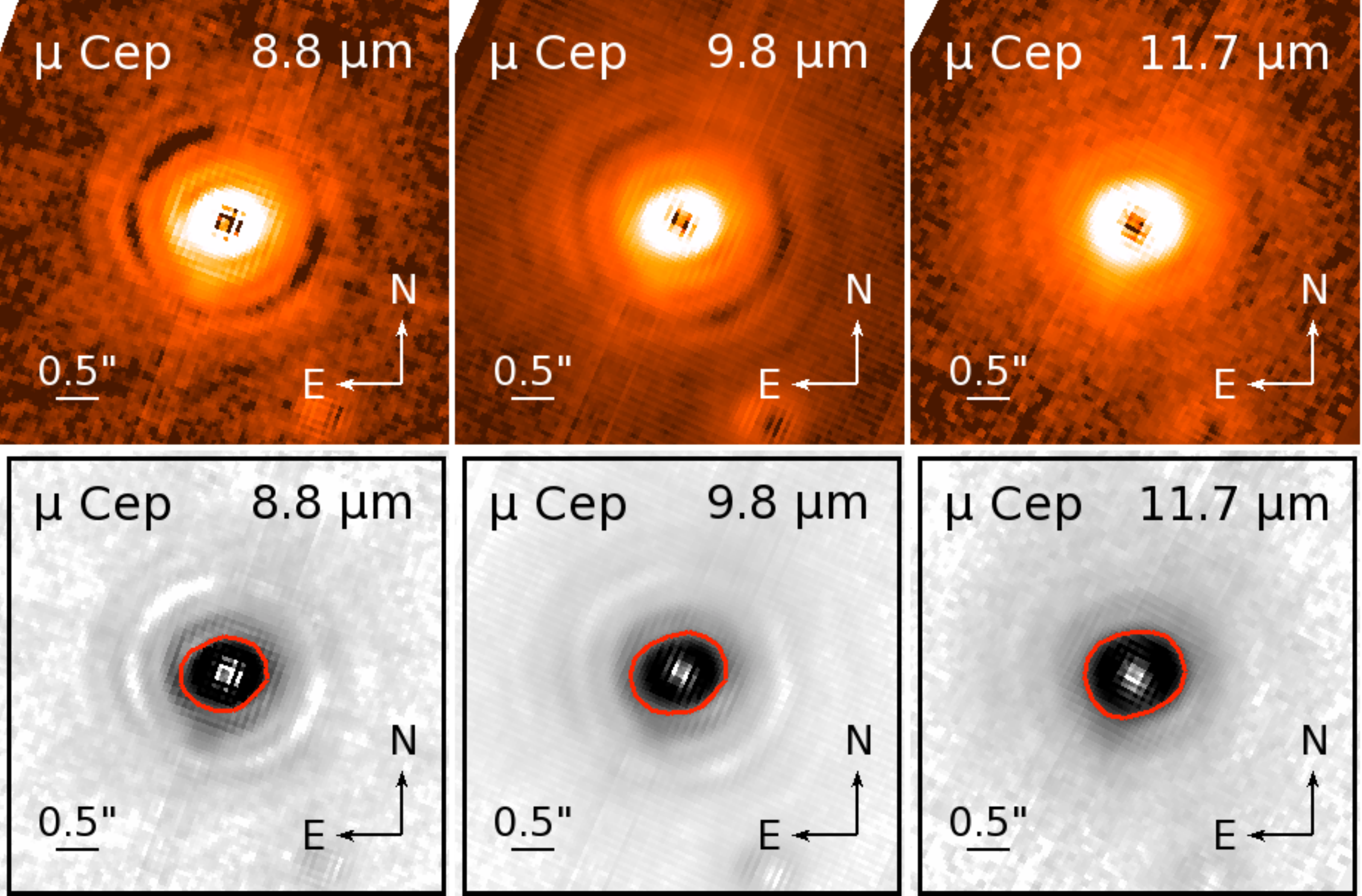}
\caption{$\mu$ Cep:   Mid-IR high resolution AO images.  \textbf{Top row:} PSF-subtracted MIRAC3 images at 8.8, 9.8 and 11.7 $\micron$ reproduced from Figure 4.3 of \citet{Schuster:2007}.  The images have been rotated to display North up, East left and the intensity is square-root scaled to emphasize extended emission.  \textbf{Bottom row:} Same images in grey-scale, each with an overlaid contour indicating an extended envelope at a position angle of $\sim$ 105$\degr$ (roughly East-West).  }
\label{muCep_MIRAC3}
\end{center}
\end{figure}

The combination of adaptive optics mid-IR imaging with MIRAC3 and wide-field far-IR imaging with SOFIA/FORCAST allow us to probe $\mu$ Cep's circumstellar environment over a broad time-scale.  We first consider its close environment.  $HST$/WFPC2 images found no evidence of scattered light within 0$\farcs$1 $-$ 0$\farcs$2 of the star \citep{Schuster:2006}.  De Wit et al. (2008) obtained ground-based diffraction-limited (0$\farcs$6) 24.5 $\micron$ images of $\mu$ Cep over multiple nights with COMICS on the 8.2 m Subaru telescope.  They detected a roughly spherical circumstellar nebula out to a radius of at least 6$\arcsec$ with an asymmetric bipolar-type geometry extending to a radius of 2$\farcs$2 from the star, oriented roughly NNE$-$SW (see their Fig. 2).  In Figure \ref{muCep_MIRAC3} we display PSF-subtracted 8 $-$ 12 $\micron$ MIRAC3 images reproduced from \citet{Schuster:2007}.  These images probe the sub-arcsecond scale closer to the star, within the region masked out in de Wit et. al's PSF-subtracted 24.5 $\micron$ image.  Schuster found an extended envelope around the star in all three filters.  The envelope has a position angle of $\sim$ 105$\degr$ East of North, which is roughly perpendicular to the 24.5 $\micron$ bipolar-type nebula.  As Schuster noted, this envelope is the likely source of $\mu$ Cep's 10 $\micron$ silicate emission feature.

\begin{figure*}
\centering
\includegraphics[scale=0.36]{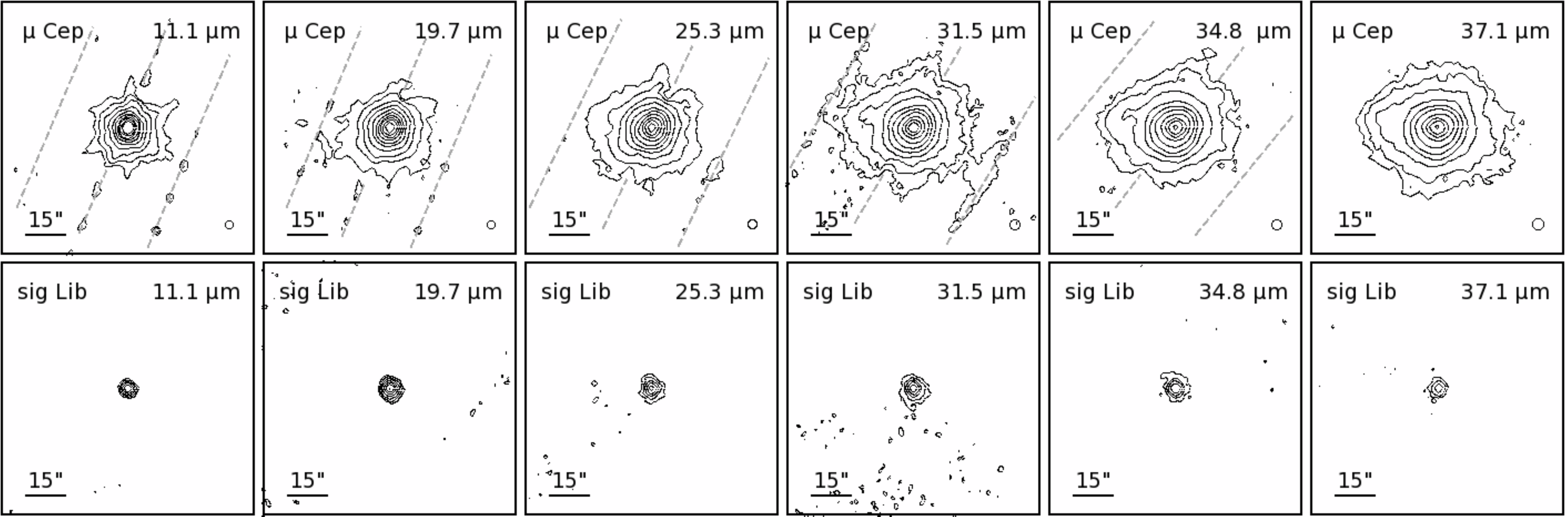}	
\caption{$\mu$ Cep:  Far-IR SOFIA/FORCAST images.  \textbf{Top row:}  Intensity contours of $\mu$ Cep in each of the FORCAST filters listed in Table 1, with central wavelengths indicated in each panel.  Each f.o.v. is $90\arcsec \times 90\arcsec$ with North up, East left.  The lowest contour is at 3-$\sigma$ above background, with each contour a factor of 2 above the preceding contour.  From shortest to longest wavelength the surface brightness of the 3-$\sigma$ contour is 45.9, 19.4, 7.4, 3.0, 2.8, and 1.9 $\times$ 10$^{-20}$ W cm$^{-2}$ arcsec$^{-2}$, respectively.  The circle in the lower right of each frame represents the beam-size (FWHM of PSF) in that filter for that flight.  The grey dashed lines in each of the 11.1 $-$ 34.8 $\micron$ images indicates horizontal on the array at the time of observation in that filter.  Cross-talk from the negative images in the chop-nod pairs caused faint enhancement and/or suppression of intensity along these lines.  It is mitigated but not completely removed by the data reduction pipeline.  The enhancements/suppressions along these lines are distinguishable as point-like artifacts smaller than the beam size.   \textbf{Bottom row:}  Standard star $\sigma$ Lib observed on the same flight (\#167), with lowest contour at 3-$\sigma$ above background, with each contour a factor of 2 above the preceding contour.  \\} 
\label{muCep_contours}
\end{figure*}

\begin{figure}
\centering
\includegraphics[scale=0.4]{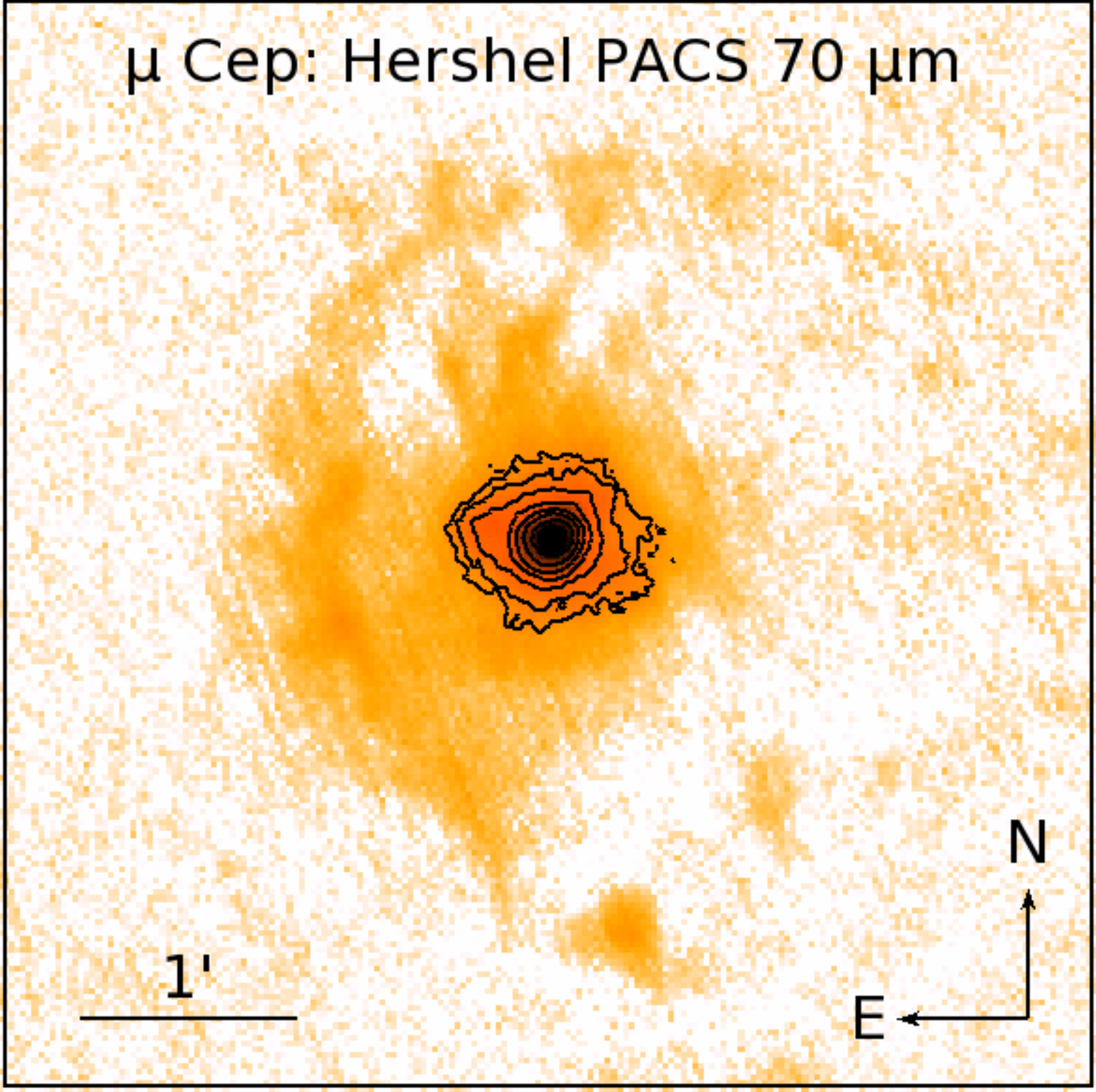}	
\caption{$\mu$ Cep:  \emph{Herschel} PACS 70 $\micron$ image showing its extensive circumstellar nebula \citep[previously published in][]{Cox:2012}.  The overlaid contours are from the SOFIA/FORCAST 37.1 $\micron$ image (see previous figure), which show the same preferential extension towards the East.\\} 
\label{muCep_overlay}
\end{figure}

On the arcminute scale, we can trace $\mu$ Cep's earlier mass loss detected in cooler dust.   Our SOFIA/FORCAST images in Figure \ref{muCep_contours} show extended emission out to a radius of 20$\arcsec$ around the star, with a particularly noticeable extension to the East out to about 25$\arcsec$.  This extension is in the same direction as the larger nebula around $\mu$ Cep seen in the \emph{Herschel} PACS images at 70 and 160 $\micron$ presented by \citet{Cox:2012}.  In Figure \ref{muCep_overlay} we overlay contours from our 37.1 $\micron$
FORCAST image on the 70 $\micron$ image, in which the nebula extends out to about 1$\farcm$5 to the southeast.  Cox et al. analyzed this nebula as a bow-shock caused by $\mu$ Cep's stellar wind interacting with the ISM.  Given the traditional assignment of $\mu$ Cep to membership in the Cep OB2 association \citep{Humphreys:1978}, an alternate possibility is that its wind may be shaped by the UV flux from  nearby OB stars, as for example is the case with the bean-shaped circumstellar nebula of NML Cyg \citep{Schuster:2009}.  However, de Zeeuw et al.'s (1999) analysis of \emph{Hipparcos} proper motions in the region containing Cep OB2 excluded $\mu$ Cep from membership in the association. The extent of the nebula indicates that $\mu$ Cep has lost mass over at least the past several thousand years.  

\subsubsection{Mass-Loss Rate}
\begin{figure}
\begin{center}
\includegraphics[scale=0.21]{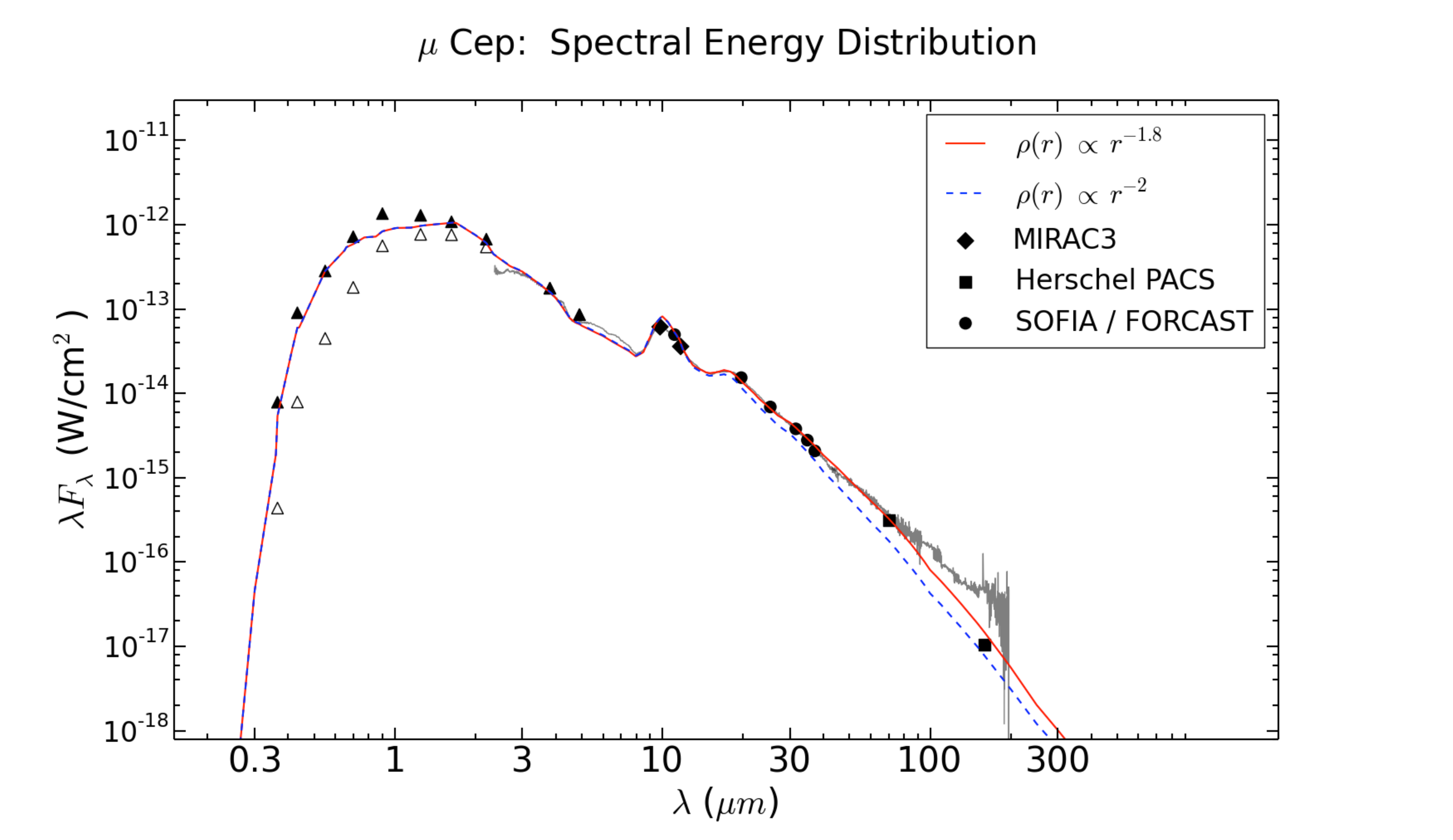}
\caption{$\mu$ Cep:  Spectral Energy Distribution.  The open triangles are observed (reddened) visual and near-IR photometry; the filled triangles are the same after dereddening for $A_{V}$ = 2.0.  The grey lines are ISO SWS and LWS spectra.  The filled diamonds are fluxes from the MIRAC3 images (this work).  The filled circles are the color-corrected SOFIA/FORCAST observations (this work), integrated out to the 3-$\sigma$ level in each image to include the extended nebular emission.  The filled squares at 70 and 160 $\micron$ are the color-corrected PACS fluxes obtained from integrating all nebular emission above the 3-$\sigma$ level.  The blue dashed line is a DUSTY model for dust density distribution $\rho(r)\propto r^{-2}$, i.e., a constant mass-loss rate $\dot{M}$.  A better fit to the observed SED is obtained for a distribution  $\rho(r)\propto r^{-1.8}$ (red solid line).}
\label{muCep_SED}
\end{center}
\end{figure}

We combine our SOFIA/FORCAST 37.1 image with DUSTY models in order to compute $\mu$ Cep's mass-loss rate.   For the stellar spectrum we use a \citet{Castelli:2004} ATLAS9 model stellar atmosphere with $T_{\star}$ = 3750 K and surface gravity log $g$ $=$ 0 appropriate for a red supergiant.  We adopt an inner shell boundary dust temperature of 1000 K, a commonly assumed sublimation temperature for silicate grains.  We assume that the outer radius of the dust shell is a factor of $10^{3}$ larger than the inner radius.  This temperature and size produce a shell extending out to approximately 2$\arcmin$, consistent with the extent of the nebula seen in the PACS 70 $\micron$ image.  We explored DUSTY models that (1) reproduce $\mu$ Cep's observed spectral energy distribution (SED) and (2) predict an average intensity profile at 37.1 $\micron$ that follows the observed profile when convolved with the FORCAST PSF.  For fitting $\mu$ Cep's SED in Figure \ref{muCep_SED} we combine visual \& near-IR photometry from \citet{Lee:1970} and \citet{Heske:1990}, which we de-redden assuming an interstellar extinction of $A_{V}$ = 2.0 \citep{Neckel:1980, Levesque:2005, Rowles:2009}.  We also plot the ISO SWS \citep{de-Graauw:1996} spectrum (\citet{Justtanont:1997}, Obs ID 08001274) and ISO LWS spectrum (Obs ID 22002005, PI: M. Barlow).  For convolving the DUSTY model intensity profile in order to compare it to the observed average 37.1 $\micron$ profile we construct a convolution kernel from observations of the asteroid Juno taken on subsequent FORCAST flights during the same flight series.  This kernel better captures the wings of the FORCAST PSF at 37.1 micron than the observations of standard star $\sigma$ Lib, due to the asteroid's higher flux in the far-IR.  This helps ensure that we do not unnecessarily attribute the broad observed profile of $\mu$ Cep to enhanced mass-loss merely through underestimating the light profile of the star.

\begin{figure}
\centering
 \includegraphics[scale=0.29]{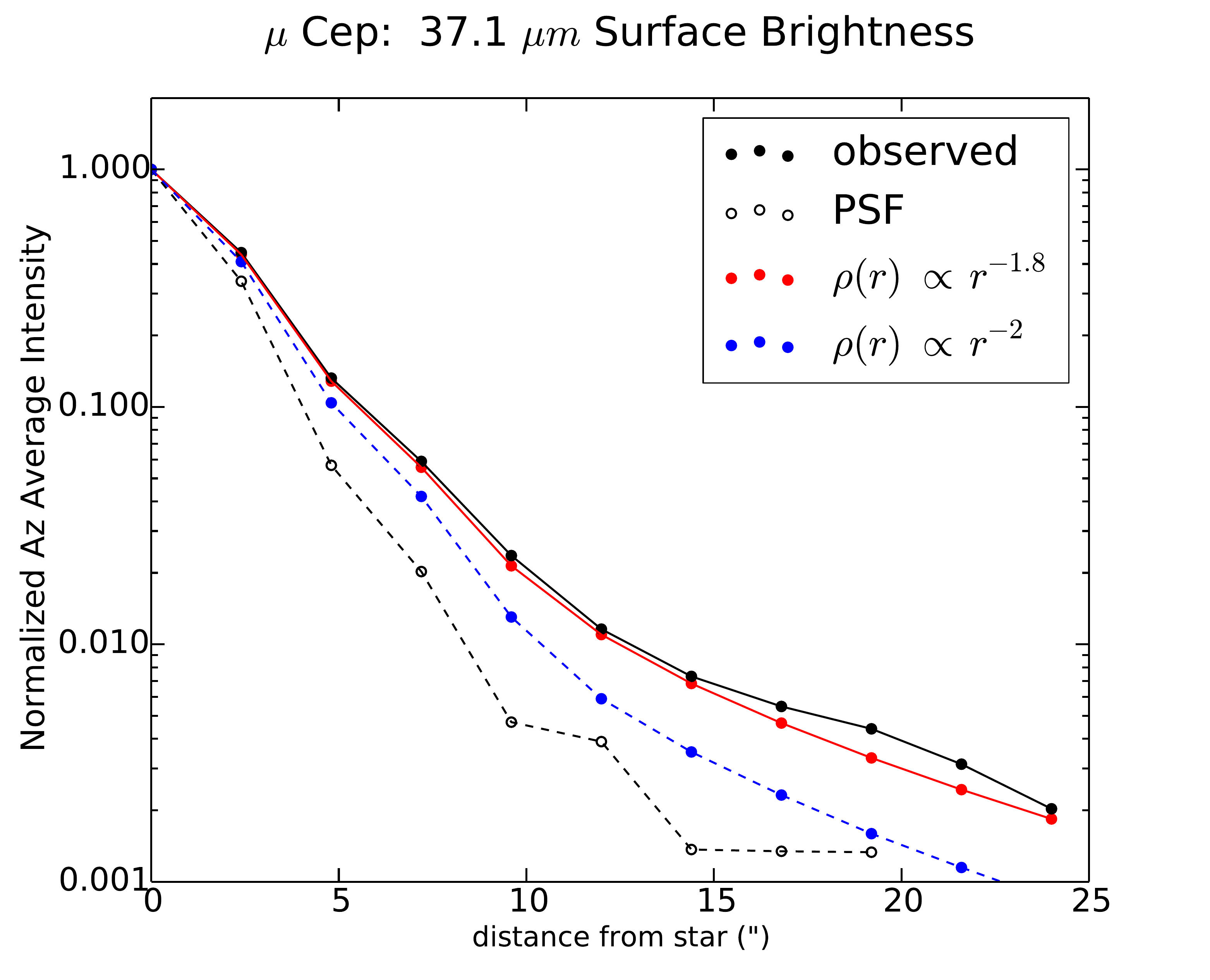} 
\caption{$\mu$ Cep:  Comparison of DUSTY model intensity profiles after convolution with PSF (black dashed line).  The solid black line is the observed azimuthal average intensity in the SOFIA/FORCAST 37.1 $\micron$ filter.  A $\rho(r)$ $\propto$ $r^{-2}$ density distribution (blue dashed line) does not predict sufficient intensity to match the observed profile.  A better fit is obtained assuming a dust density distribution $\rho(r)$ $\propto$ $r^{-1.8}$ (red solid line).  For the latter case the average mass-loss rate over the 13,000 yr age of the shell is $\langle\dot{M}\rangle$ $\approx$ 4 $\times$ 10$^{-6}$ $M_{\sun}$ yr$^{-1}$.  }
\label{muCep_F371profile}
\end{figure}

For the dust optical properties we use the ``warm'' circumstellar silicates from \citet{Ossenkopf:1992}, and assume the grain radii follow an MRN size distribution $n(a)\propto a^{-3.5}~da$ \citep{Mathis:1977} with $a_{min}=0.005~\micron$ and $a_{max}=0.25~\micron$.   We assume a power law dust mass density distribution $\rho(r) \propto r^{-q}$ with a single index $q$ throughout the shell.   We scale model SEDs to match $\mu$ Cep's dereddened visual flux, adjusting the optical depth to achieve a satisfactory fit to the SED across the mid-to-far IR range while also reproducing the observed average intensity profile at 37.1 $\micron$.  We explored a range of optical depths and a range of power law indices $q$ $\leq$ 2.  A power law index of $q$ $=$ 2 fits the SED very well through $\lambda$ $\approx$ 12 $\micron$, but produces insufficient cooler dust to explain the long wavelength end of the observed SED (see Figure \ref{muCep_SED},  blue dashed line).  Moreover, the $q$ $=$ 2 case fails to adequately reproduce the broad radial profile of the 37.1 micron intensity image (see Figure \ref{muCep_F371profile}, blue dashed line).   We find that a density distribution of $\rho(r)$ $\propto$ $r^{-1.8}$ provides both a much better fit to the far-IR SED and the observed radial profile of the intensity (red solid lines in Figures \ref{muCep_SED} \& \ref{muCep_F371profile}).  

Adopting an expansion velocity of $v_{exp}$ $=$ 35 km s$^{-1}$ from the width of its CO lines \citep{De-Beck:2010} yields a dynamical age of 13,000 yr for this shell model.  If we assume this velocity is constant throughout the shell, then per \S{3.1} above for this best-fit case with density power law index $q$ $=$ 1.8 the mass-loss rate $\dot{M}(t)$ has \emph{decreased} from 5 $\times$ 10$^{-6}$ $M_{\sun}$ yr$^{-1}$ to $\sim$ 1 $\times$ 10$^{-6}$ $M_{\sun}$ yr$^{-1}$ over this time period.  Alternately, if we simply integrate the density distribution throughout the shell to compute a total mass and divide by this age for the shell, we obtain an average mass-loss rate of $\langle\dot{M}\rangle$ $\approx$ 4 $\times$ 10$^{-6}$ $M_{\sun}$ yr$^{-1}$.  This rate exceeds the 9 $\times$ $10^{-7}$ $M_{\sun}$ yr$^{-1}$ found from its IRAS 60 $\micron$ excess \citep{Jura:1990a}, and the $4.5~\times~10^{-7}~M_{\sun}$ yr$^{-1}$ found by \citet{de-Wit:2008}, though it is in good agreement with the $2 ~\times ~10^{-6} ~M_{\sun}$ yr$^{-1}$ found by De Beck et al. from modeling its CO line emission.  Our rate is about half that of the $10^{-5}M_{\sun}$ yr$^{-1}$ found by \citet{Gehrz:1971} based on modeling its 11 $\micron$ emission, with the difference most likely attributable to our model using a gas-to-dust ratio that is 40\% of the value used in that work and/or differences in the grain opacity used.   In general all investigations of $\mu$ Cep's mass-loss rate using various methods have found it is lower than average compared to other high-luminosity RSGs \citep[see e.g., Figure 3 of][]{Mauron:2011}.

\subsection{VY CMa}
Multi-wavelength \emph{HST} visual imaging of the red hypergiant VY CMa has revealed extensive, asymmetric mass-loss within the past 500 - 1000 years, dominated by multiple spatially and kinematically distinct features, indicating independent ejection events from highly localized regions of the star \citep{Smith:2001, Humphreys:2005,Humphreys:2007}.  Three of the most prominent ejections, the Northwest Arc, Arc 1 and Arc 2 \citep[see Figures 1 $-$ 4 of][]{Humphreys:2007}, each contain roughly $3\times10^{-3}$ $M_{\sun}$ of gas and dust, which accounts for about 10\% of the total ejected mass of 0.2 $-$ 0.4 $M_{\sun}$ in the nebula out to a radius of 6$\arcsec$.  At the same time, Keck/HIRES spectra showed that the more diffuse uniformly distributed gas and dust is surprisingly stationary, with little or no velocity relative to the star.  On this basis \citet{Humphreys:2005} argued that the high mass loss rate of $\sim$10$^{-4}$ $M_{\sun}$ previously found for VY CMa is an average measure of the mass carried out by these specific ejections accompanied by streams or flows of gas through low-density regions in the dust envelope.  The unusually diverse molecular chemistry of VY CMa's circumstellar envelope appears to be directly related to this history of independent, localized ejections, with certain species of molecules associated with the distinct blue-shifted and red-shifted features in the ejecta \citep{Ziurys:2007, De-Beck:2015}.  The star's recent history continues to show strong, discrete mass loss events.  High resolution near-IR imaging and  polarimetry found the more recently ejected ``Southwest Clump'' feature ($\sim$ 1$\farcs$5 from the star) to be optically thick to scattering out to 5 $\micron$ with a total mass (gas + dust) of at least $5~\times~10^{-3} ~M_{\sun}$ \citep{Shenoy:2013, Shenoy:2015}.  Most recently, ALMA sub-millimeter observations resolved prominent dust components even closer to the star, with the brightest component (dubbed component ``C'') having a total mass of gas and dust of at least 2.5 $\times~10^{-4}~M_{\sun}$ \citep{OGorman:2015}.  

\begin{figure*}
\centering
\includegraphics[scale=0.37]{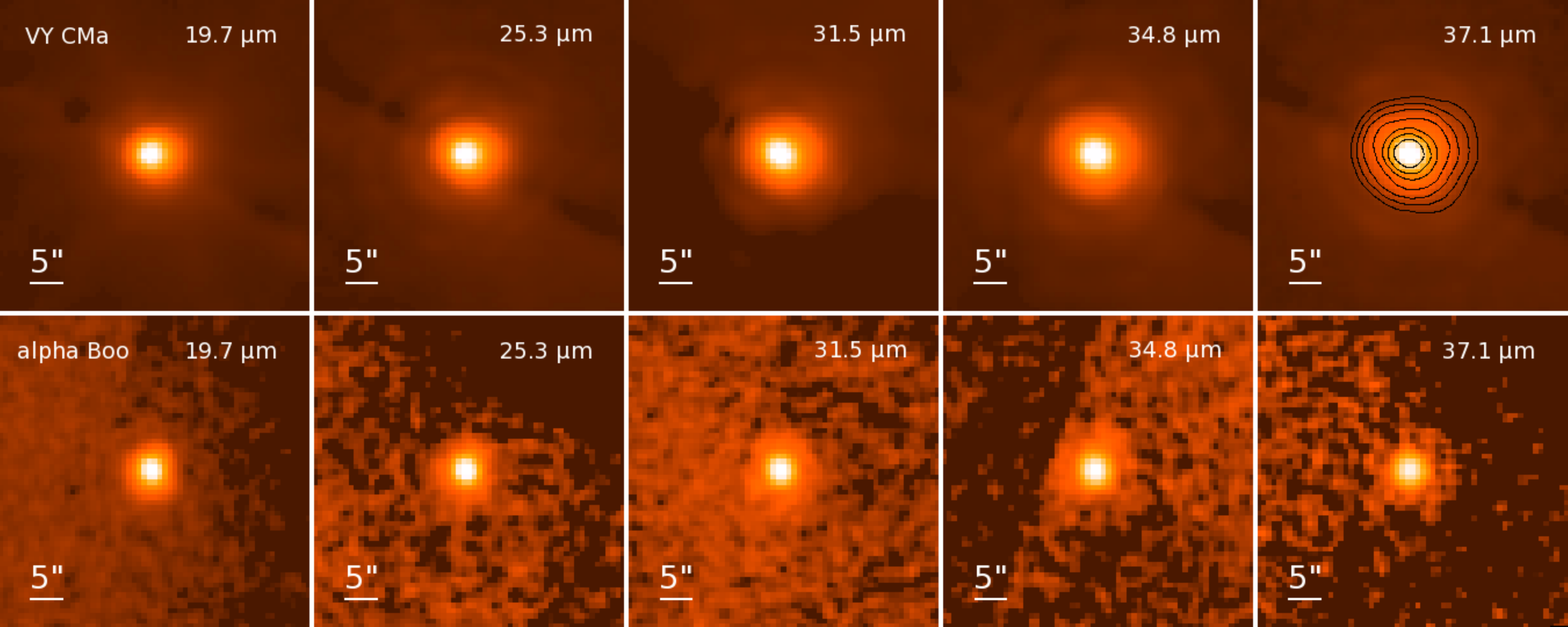}	
\caption{VY CMa:  Far-IR SOFIA/FORCAST images.  \textbf{Top row:}  VY CMa in each of the FORCAST filters listed in Table 1, with central wavelengths indicated in each panel.  Each f.o.v. is $45\arcsec \times 45\arcsec$ with North up, East left.  The intensity is square-root scaled.  \textbf{Bottom row:}  Standard star $\alpha$ Boo observed on the same flight (\#154).  The contours on the 37.1 $\micron$ image of VY CMa are the same contours overlaid on the \emph{HST} image in Figure \ref{VY_overlay}.\\} 
\label{VY_images}
\end{figure*} 

\begin{figure}
\centering
\includegraphics[scale=0.3]{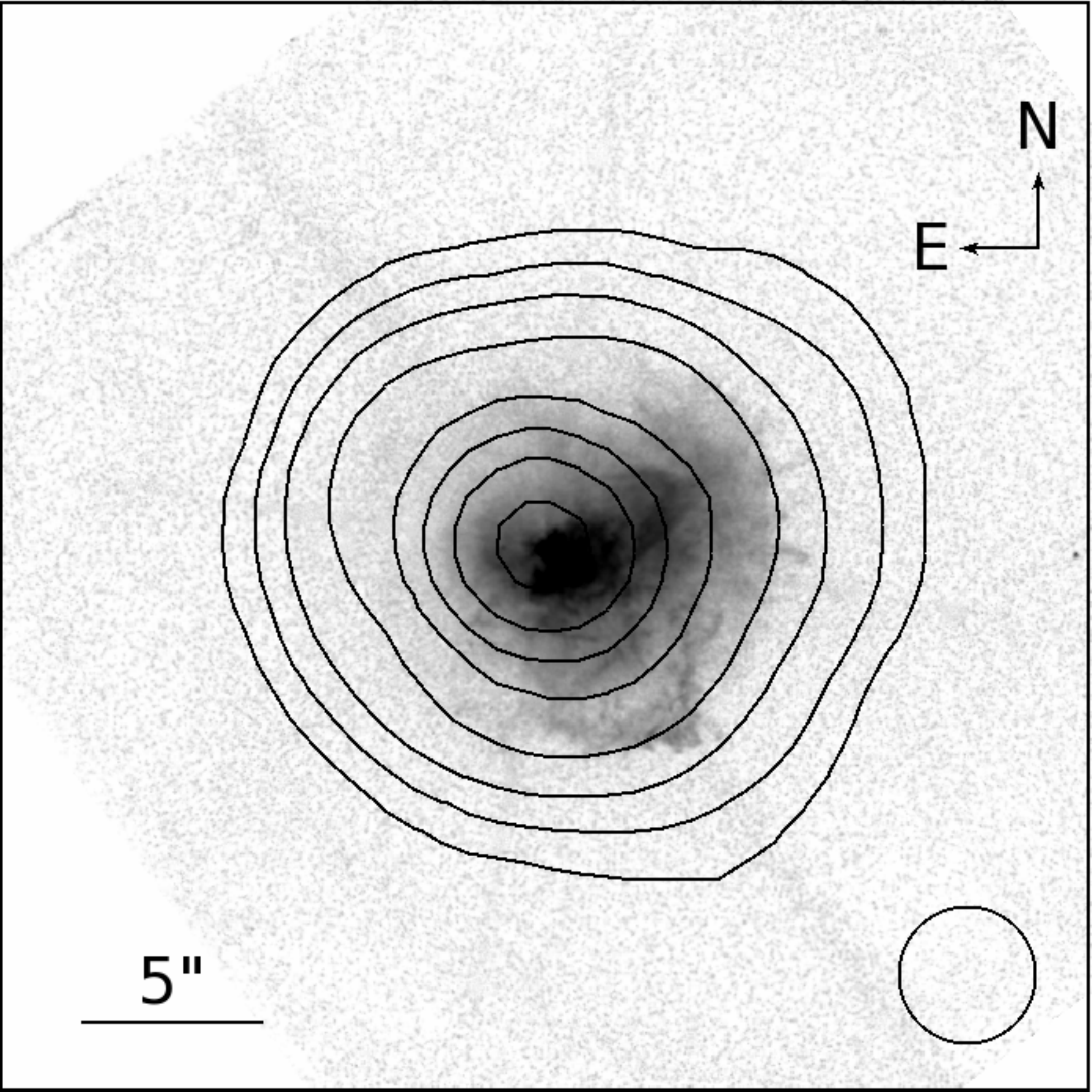} 	
\caption{VY CMa:  FORCAST 37.1 $\micron$ total intensity contours from top right of Figure \ref{VY_images} overlaid on $HST$ F547M ($\lambda_{0}$ = 0.55 $\micron$) visual image reproduced from \citet{Smith:2001}.  The f.o.v. is $30\arcsec\times 30\arcsec$ with North up, East left.  The circle in the lower right represents the FORCAST 37.1 $\micron$ beam size (PSF FWHM = 3$\farcs$7).  The extension towards both the northwest and southwest indicates cooler dust not detectable in scattered light.  The extensions in these directions correspond to the directions of ejection of the features respectively named as the NW Arc and Arc 1 by \citet{Humphreys:2007} (see their Fig. 3, the second-epoch $HST$ F547M image).\\} 
\label{VY_overlay}
\end{figure}

VY CMa's appearance in the SOFIA/FORCAST images (Figure \ref{VY_images}) is broader than a point-source, although in contrast to $\mu$ Cep we find no evidence of a long history of mass loss.  Interestingly, the appearance of the FORCAST image follows the general shape seen in the visual.  In Figure \ref{VY_overlay} we show intensity contours from the FORCAST 37.1 $\micron$ image overlaid on the \emph{HST} visual image from \citet{Smith:2001}.   The FORCAST images are extended towards the northwest and southwest in the same directions as the prominent Northwest Arc and Arc 1 seen in the scattered visual light, which were ejected $\sim$ 400 and 900 yr ago respectively.  The NW Arc and Arc 1 are expanding at 40 and 50 km s$^{-1}$ respectively relative to the star, into uniformly distributed gas and dust that is either stationary or only slowly expanding away \citep{Humphreys:2005}.  The extension of the longer wavelength FORCAST images in the same directions may indicate that cooler dust lost earlier is being swept up by these more recent, fast moving ejections.  The 37.1 $\micron$ image is extended in the east and northeast directions as well, in the direction of a dark cloud where the extinction is especially high and is responsible for its asymmetric shape in the visual \citep[see Figure 9 of][]{Smith:2001, Montez:2015}.  Thermal emission at 37.1 $\micron$ is able to penetrate this obscuring dust.

\begin{figure}
\centering
\includegraphics[scale=0.22]{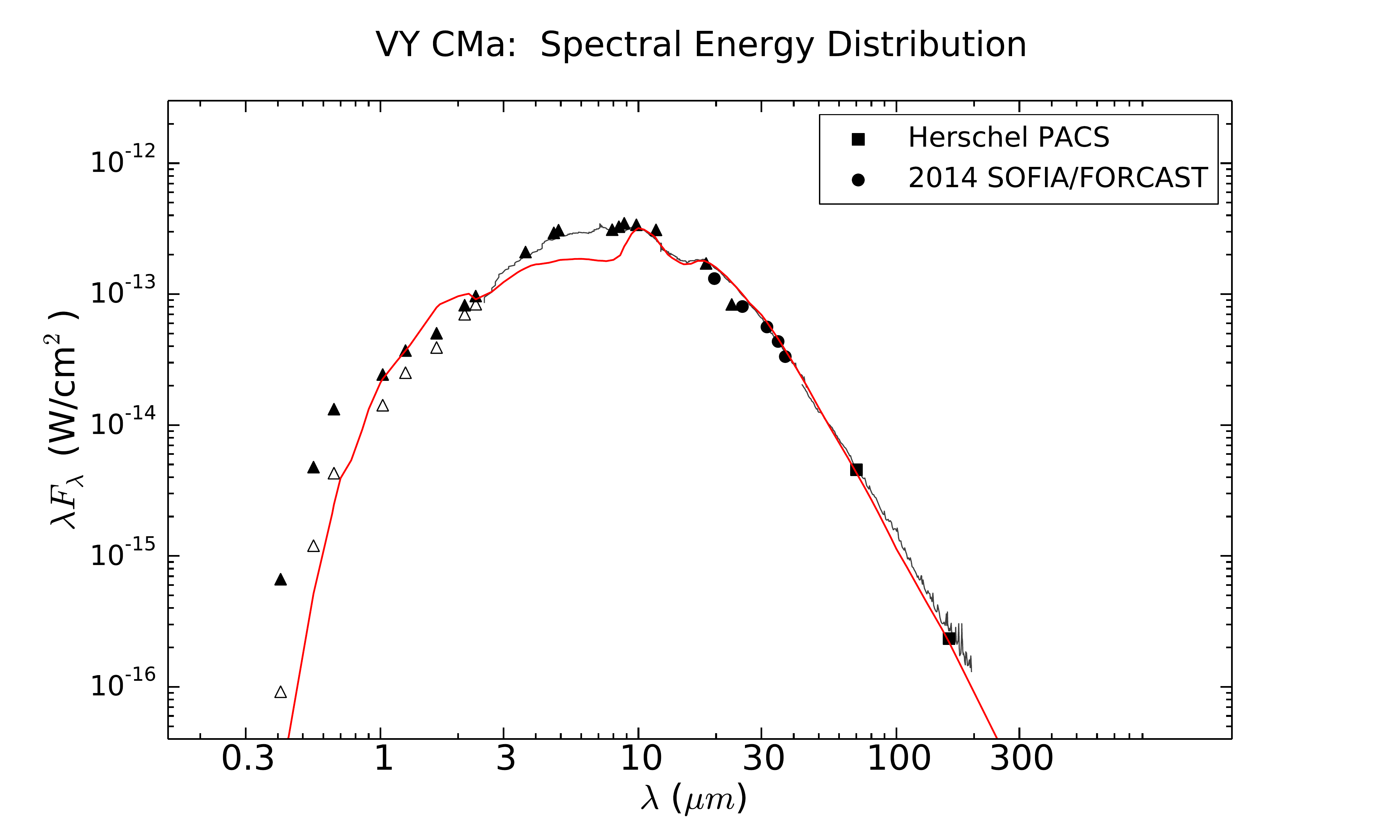} 
\caption{VY CMa:  Spectral Energy Distribution.  The triangles are photometry from \citet{Smith:2001}, with the open triangles showing the observed values prior to de-reddening for $A_{V}$ $=$ 1.5.  The grey solid lines are the ISO-SWS spectrum \citep{Harwit:2001} and the ISO-LWS spectrum \citep{Polehampton:2010}.  The black circles are the color-corrected fluxes measured from the SOFIA/FORCAST observations using an aperture of radius = 12$\arcsec$.  The 6\% uncertainty from the flux calibration is smaller than the plotted symbol.   The black squares at 70 and 160 $\micron$ are the color-corrected fluxes from the PACS images.  The red solid line is the DUSTY model fitted to the SED, yielding an average mass-loss rate of $\langle\dot{M}\rangle$ $\approx$ 6 $\times$ 10$^{-4}$ $M_{\sun}$ yr$^{-1}$.}
\label{VY_SED}
\end{figure}

We estimate VY CMa's mass-loss rate with an elementary DUSTY model.  Given VY CMa's highly asymmetric nebula seen in the optical and near-IR, the assumption of spherical symmetry is obviously a simplification.  Previous studies have applied radiative transfer models that include a circumstellar disk and a variety of viewing angles, and/or a range of mineralogies for the dust, with a thorough exploration presented in \citet{Harwit:2001}.  At the resolution of our FORCAST images, however, spherical symmetry is an acceptable assumption that should provide an accurate average mass-loss rate.  As seen in Figure \ref{VY_SED}, VY CMa's SED is highly reddened.  The interstellar reddening towards it is estimated at no more than $A_{V}$ $=$ 1.5, the value we have used to de-redden the optical and near-IR photometry (open triangles).  It is thus clear that the very red energy distribution is mostly due to circumstellar extinction \citep[see also Appendix of][]{Humphreys:2007}.  For our DUSTY model we use an ATLAS9 stellar atmosphere with $T_{\star}$ $=$ 3500 K \citep{Wittkowski:2012}.  We find that we cannot fit the FORCAST and PACS fluxes with any model using silicate grains alone.  We are able to fit the long-wavelength end of its spectrum if we assume a grain mixture that is 50-50 silicates and metallic Fe, with the same iron optical constants used by  \citet{Harwit:2001}.  We use a grain size distribution $\propto$ $a^{-3.5}$, though we find that to account for the broad, rising spectrum beyond the near-IR we require large grains capable of scattering in the 2 $-$ 5 $\micron$ range.  We  use an upper-end grain radius of $a_{max}$ $=$ 5 $\micron$.  Increasing the maximum grain radius from the submicron scale has minimal impact on our computed mass-loss rate since the dominant contribution to the opacity comes from the smallest grains in the distribution.  We first attempted to fit the SED assuming a constant mass-loss rate density distribution of $\rho(r)$ $\propto$ $r^{-2}$ but found it inadequate to explain the far-IR flux.  Our best-fit model displayed in Figure \ref{VY_SED} assumes $\rho(r)$ $\propto$ $r^{-1.5}$ throughout a shell with an outer radius of 10$\arcsec$.  Although it underpredicts the visual fluxes and strength of the 3 $-$ 8 $\micron$ region relative to the 10 $\micron$ silicate feature, our model fits reasonably well to the JHK range where the effect of ISM reddening is the least uncertain and provides a very good fit to the FORCAST and PACS fluxes.

For this best-fit model the optical depth $\tau$(37.1 $\micron$) $=$ 0.17.  Assuming a 100:1 gas-to-dust mass ratio, we compute a total mass of 0.7 $M_{\sun}$ in the nebula.  To determine the average mass-loss rate, we estimate the age of the 10$\arcsec$ radius model shell using a constant expansion velocity.  Assuming a constant expansion velocity is an approximation, since the Keck/HIRES spectra showed that a single uniform expansion velocity is not found throughout the nebula \citep{Humphreys:2005}.  That work found the large arcs to the northwest and southwest expanding at 40 $-$ 50 km s$^{-1}$, while the more uniformly distributed gas showed little or no motion.   Very close to the star ($<$ 0$\farcs$4), the proper motion of H$_{2}$O masers indicates an expansion velocity of 27 km s$^{-1}$ per \citet{Richards:1998} \citep[rescaled for $D$ = 1.2 kpc per][]{Zhang:2012}.  \citet{De-Beck:2010} found an expansion velocity of $\sim$ 47 km s$^{-1}$ based on the widths of CO lines in the circumstellar envelope.  We adopt this last value as an average velocity, since CO molecules are found throughout the circumstellar envelope.  This yields an age of $\Delta t$ = 1200 yr for the 10$\arcsec$ radius model shell.  The resulting average mass-loss rate is $\langle\dot{M}_{tot}\rangle$ $\approx$ 6 $\times$ 10$^{-4}$ $M_{\sun}$ yr$^{-1}$.  Since this is an average rate, it does not capture the higher short-term rates on the order of 10$^{-3}$ $M_{\sun}$ yr$^{-1}$ associated with the NW Arc and Arcs 1 and 2 \citep{Humphreys:2005}.  Our average rate is within the same range of several $\times$ 10$^{-4}$ $M_{\sun}$ yr $^{-1}$ found by previous studies \citep{Danchi:1994,Knapp:1985a,De-Beck:2010}.

\subsection{IRC +10420}
\subsubsection{Previous Visual, Near-IR and Sub-Millimeter Studies}
IRC +10420 is one of a few intermediate temperature stars defining the empirical upper luminosity boundary in the H-R Diagram.  With $L \sim 5\times10^{-5}L_{\sun}$, its  A-F type spectrum and extensive visible ejecta  make it a special example of post-RSG evolution \citep{Humphreys:2002}.  \citet{Jones:1993} used visual polarimetry to determine that most of the extinction towards IRC +10420 is interstellar; combining the polarimetry with the reddening of its optical SED and radial velocity they established a distance of $\sim$5 kpc.  $HST$ visual imaging reveal a very complex multi-stage mass loss history, with one or more reflection shells at $5\arcsec-6\arcsec$ that were probably ejected during the star's previous RSG stage \citep{Humphreys:1997}, while the near environment within a radius of 2$\arcsec$ shows many jet-like structures, rays and arcs  in the ejecta.  For radii from 2$\arcsec$ to 5$\arcsec$, Humphreys et al. used the surface brightness of the visual scattered light to estimate $\dot{M} \sim 2.4\times10^{-4}M_{\sun}$ yr$^{-1}$.  Most interestingly, they found evidence of enhanced mass-loss at a rate of $\sim 10^{-3} M_{\sun}$ yr$^{-1}$ based on the near-IR scattered light within 1$\arcsec$ of the star, suggesting a very high mass loss stage within the past 400 years that has recently ceased.  Second epoch $HST$/WFPC2 imaging combined with $HST$/STIS long-slit spectroscopy showed that the many numerous arcs, knots and condensations in the inner region were ejected at different times in different directions, but were all nearly in the plane of the sky \citep{Tiffany:2010}.   Thus we view IRC +10420 nearly pole-on.  This geometry has been confirmed  by interferometry of Br$\gamma$ emission from neutral and ionized gas  around the star \citep{Oudmaijer:2013} and  by high-resolution adaptive optics 2.2 $\micron$ imaging polarimetry \citep{Shenoy:2015}.

\subsubsection{Mass-Loss History: Mid-to-Far-IR Imaging}
\begin{figure}
\centering
\includegraphics[scale=0.27]{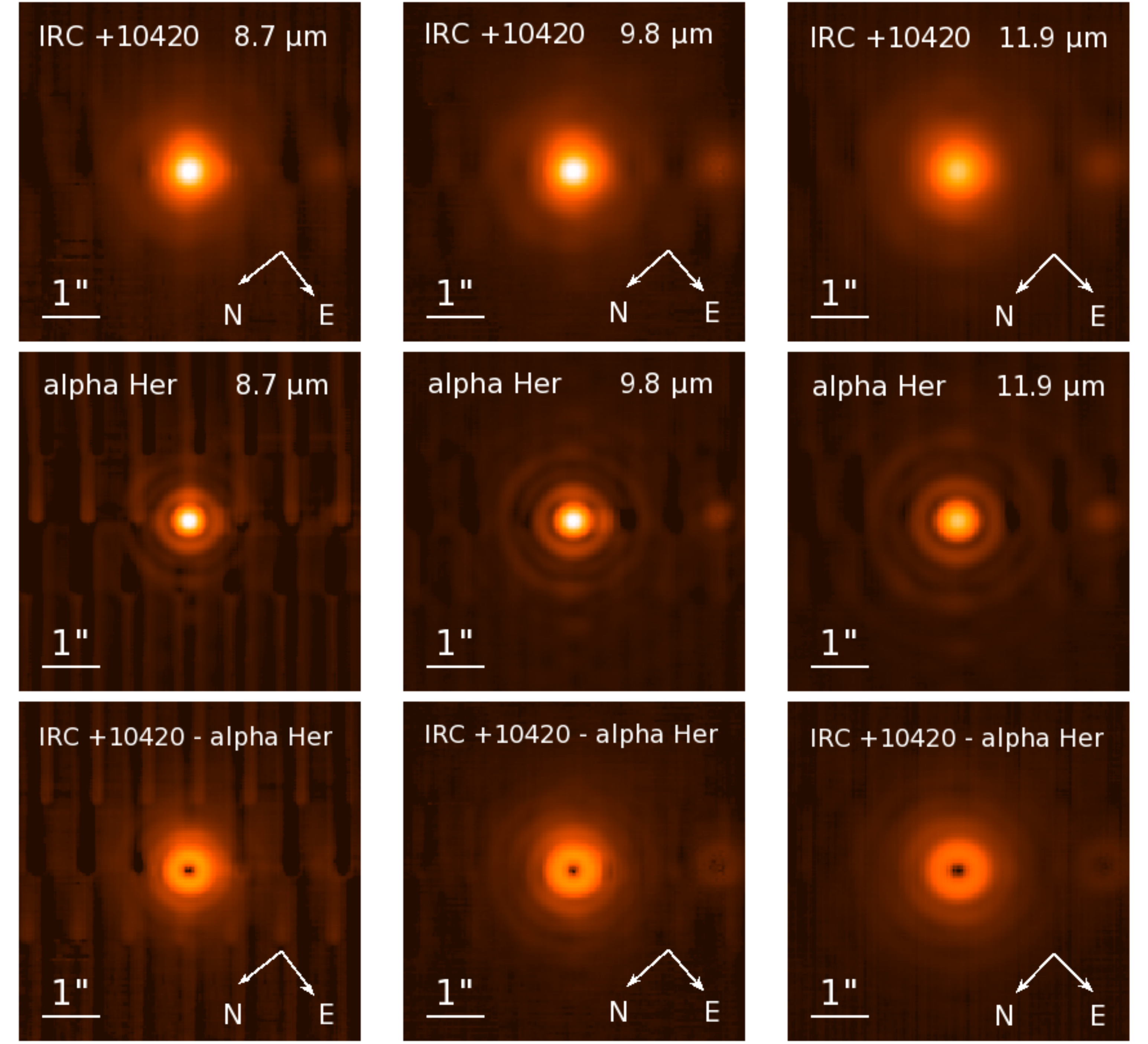}
\caption{IRC +10420:  Mid-IR high resolution images.  \textbf{Top row:}  MMT/MIRAC4 adaptive optics thermal infrared images of IRC +10420 at 8.7, 9.8 and 11.9 $\micron$ taken on UT 2008 Jun 16.  The intensity is square-root scaled.   \textbf{Middle row:}  Same, for the PSF star $\alpha$ Her observed immediately after IRC +10420 at similar airmass.  The first minima in the Airy ring patterns occur at radii of 0$\farcs$40, 0$\farcs$45, and 0$\farcs$53 in the 8.7, 9.8 and 11.9 $\micron$ filters respectively.  For comparison the diffraction limits (1.22$\lambda/D$) are 0$\farcs$34, 0$\farcs$38, and 0$\farcs$46 respectively at those wavelengths for the 6.5-m MMT aperture.  \textbf{Bottom row:}  IRC +10420 after subtraction of the scaled PSF image, stretched to the same maximum intensity as the corresponding top row image.  A clear excess of extended emission around the star is present out to a radius of $\sim2\arcsec$ in all three filters.} 
\label{IRC_MIRAC4}
\end{figure}

IRC +10420 has one of the strongest 10 $\micron$ silicate emission features among the hypergiants.   With MIRAC4 we can probe the sub-arcsecond scales where this emission originates.  In Figure \ref{IRC_MIRAC4} we show the MIRAC4 8.7, 9.8 and 11.9 $\micron$ images of IRC +10420 along with comparisons to the PSF star $\alpha$ Her.  Subtraction of the PSF leaves circularly symmetric extended emission out to a radius of $\sim1\farcs5$  in all three filters.  The extended emission does not show the NE-SW preference previously seen at the same wavelengths with similar resolution in \citet{Humphreys:1997} and \citet{Meixner:1999}.   It is likely that some of the strong 10 $\micron$ silicate emission arises from this extended envelope.  

\begin{figure*}
\centering
\includegraphics[scale=0.36]{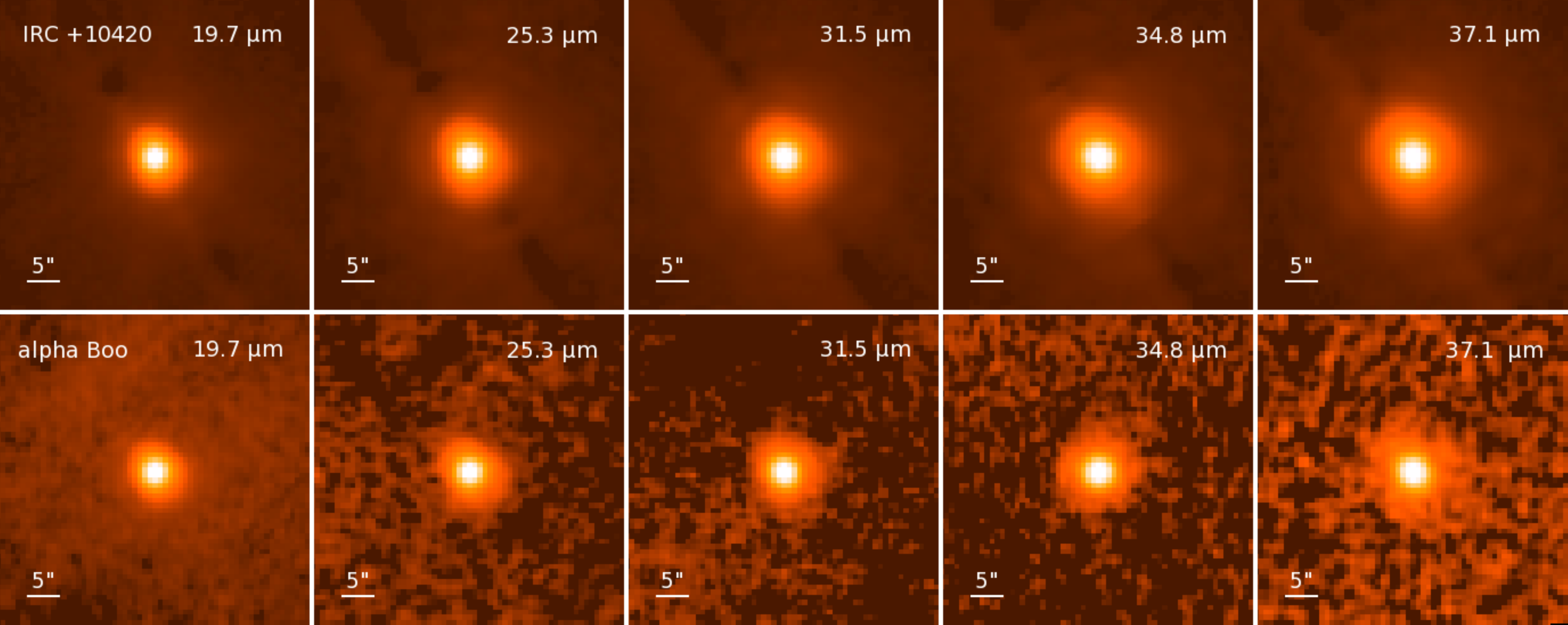}	
\caption{IRC +10420:  Far-IR SOFIA/FORCAST images.  \textbf{Top row:}  IRC +10420 in each of the FORCAST filters listed in Table 1, with central wavelengths indicated in each panel.  Each f.o.v. is $45\arcsec \times 45\arcsec$ with North up, East left.   The intensity is square-root scaled. \textbf{Bottom row:}  Standard star $\alpha$ Boo observed on the same flight (\#177).\\ \\} 
\label{IRC_contours}
\end{figure*}

\begin{figure}
\centering
\includegraphics[scale=0.37]{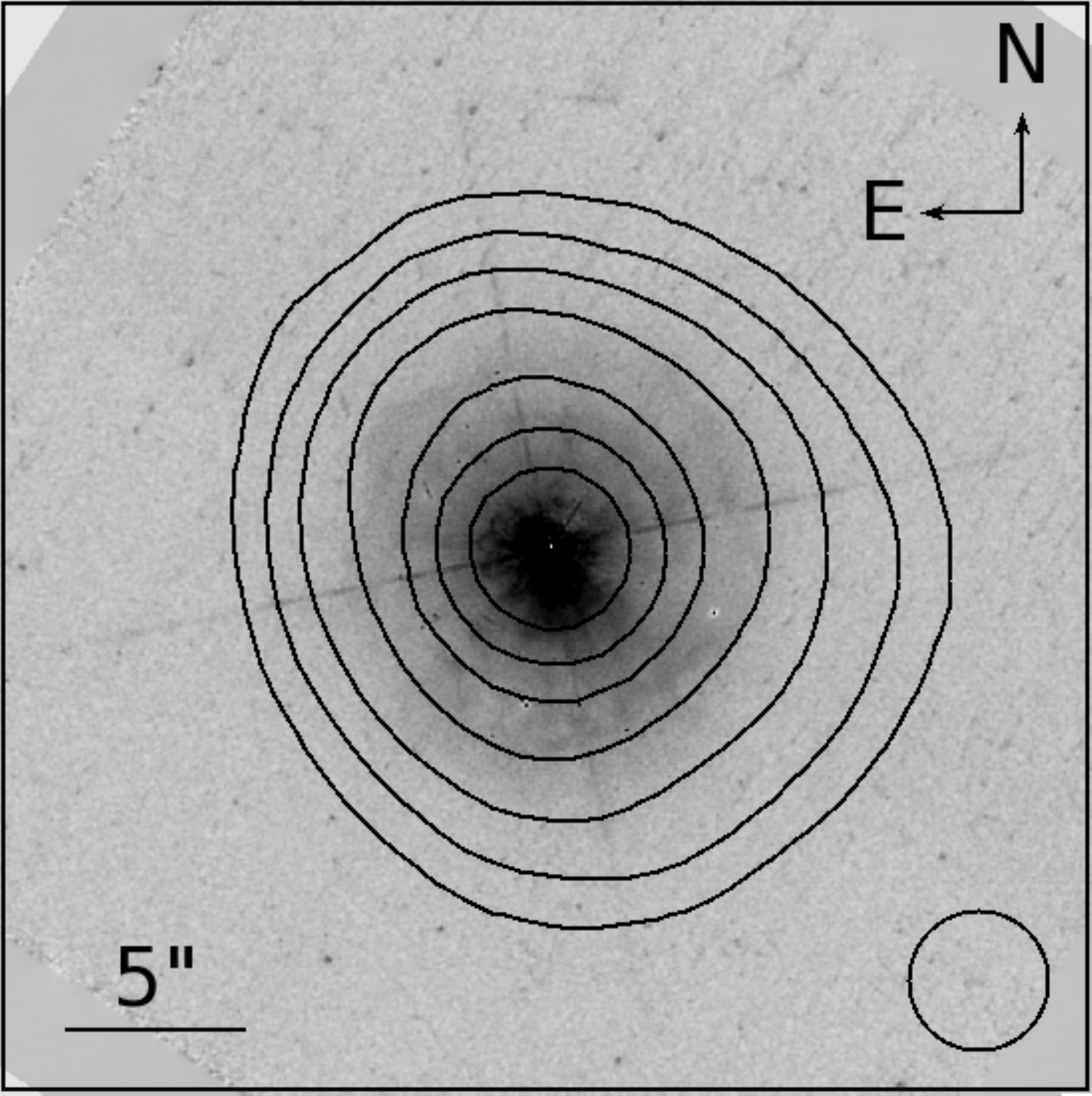}	
\caption{IRC +10420 37.1 $\micron$ total intensity contours from top right of Figure \ref{IRC_contours} overlaid on $HST$ F547M ($\lambda_{0}$ = 0.55 $\micron$) visual image reproduced from \citet{Tiffany:2010}.  The f.o.v. is $30\arcsec\times 30\arcsec$ with North up, East left.  The circle in the lower right represents the FORCAST 37.1 $\micron$ beam size (PSF FWHM = 3$\farcs$9).  The extension towards the southwest coincides with the enhancement in the mass loss in that direction seen in both the visual and sub-millimeter.} 
\label{IRC_overlay}
\end{figure}

Farther  from the star, the $HST$ visual images showed the ejecta separated into distinct, approximately spherical shells at $5\arcsec-6\arcsec$ from the star.  The long-exposure $HST$ F675W image also showed evidence of more distant ejecta as much as $8\arcsec-9\arcsec$ away \citep[see Figure 5 of][]{Tiffany:2010}, which is also seen in the near-IR in J-band polarimetry \citep{Kastner:1995}.  Our FORCAST images in Figure \ref{IRC_contours} probe a similar angular extent in radius, with emission in the longest filters detected out to a radius of  $\sim$10$\arcsec$.  The $HST$ visual observations showed a pronounced enhancement of the mass-loss towards the southwest of the star.  This morphology is seen in our FORCAST images as well, as shown in Figure \ref{IRC_overlay} where we overlay the FORCAST 37.1 $\micron$ intensity contours on the $HST$ visual image from \citet{Tiffany:2010}.   Sub-millimeter observations find extension in the same direction as well \citep{Dinh-V.-Trung:2009}.

\begin{figure}
\begin{center}
\includegraphics[scale=0.21]{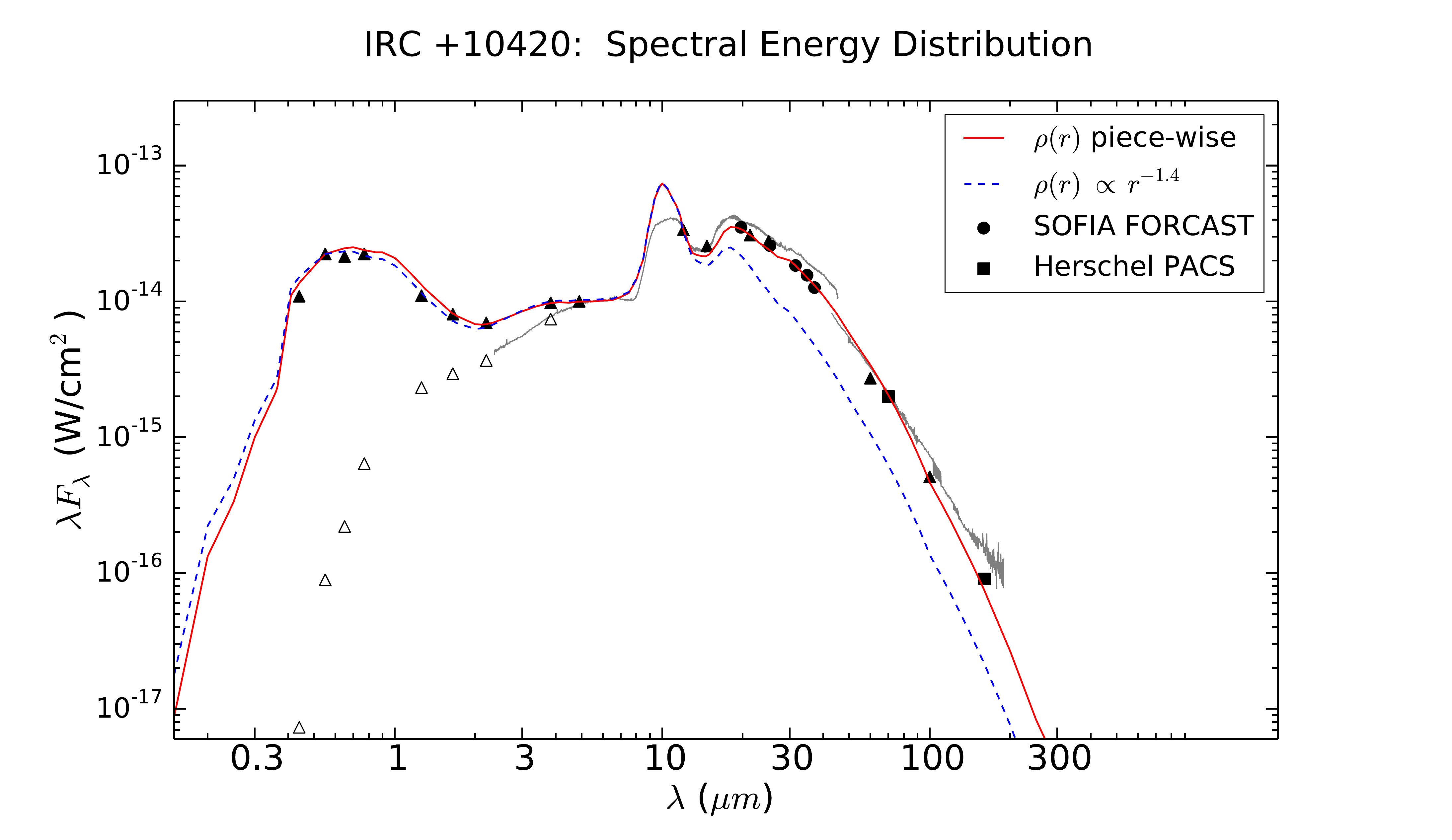}
\caption{IRC +10420:   Spectral Energy Distribution.  The black triangles are point-source photometry compiled from \citet{Jones:1993}, \citet{Oudmaijer:1996}, \citet{Humphreys:2002} and the IRAS and MSX Point Source Catalogs (with color correction).  The thin grey lines are the ISO SWS and LWS spectra.  The black circles are the color-corrected fluxes measured from the SOFIA/FORCAST observations using an aperture of radius = 10$\arcsec$.  The assumed 6\% uncertainty from the flux calibration is smaller than the plotted symbol.  The black squares at 70 and 160 $\micron$ are the color-corrected fluxes from the \emph{Herschel}/PACS images.  The blue dashed line is a DUSTY model for a density distribution $\rho(r)$ $\propto$ $r^{-1.4}$ throughout the shell, which adequately fits the SED through $\sim$ 12 $\micron$ but falls off too rapidly at longer wavelengths.  The red solid line is the best-fit model, constructed using the same power-law index but with a density enhancement as depicted in the next figure. }
\label{IRC_SED}
\end{center}
\end{figure}

To explore IRC +10420's mass-loss history we use DUSTY to model its SED, requiring the models to account for the substantial extended emission seen in the 11.9 $\micron$ MIRAC4 image out to a radius of $\sim$2$\arcsec$.  To build its SED, in Figure \ref{IRC_SED} we compile photometry from the optical through L$'$ (3.8 $\micron$) from \citet{Jones:1993}, \citet{Oudmaijer:1996} and \citet{Humphreys:2002}, which we de-redden assuming $A_{V}$ = 6.  For the mid-to-far IR we include color-corrected fluxes from the IRAS and MSX Point Source catalogs as well as the ISO SWS spectrum (Obs ID 12801311) as processed by \citet{Sloan:2003} and the ISO LWS spectrum from \citet{Lloyd:2003} (Obs ID 72400605).  To this we add the photometry from our SOFIA/FORCAST images and the \emph{Herschel}/PACS images.\footnote{\footnotesize{~The good agreement between the trend of our FORCAST photometry and the start of the LWS spectrum indicate that the discontinuity between the end of the SWS spectrum and the start of the LWS spectrum is an artifact of the \citep{Sloan:2003} reprocessing of the SWS spectrum, as previously noted by \citet{Ladjal:2010}.}}  The conventional  stellar effective temperature is ill-defined for IRC +10420 \citep{Humphreys:2002}, though a value between 8000 $-$ 8500 K is likely \citep[see e.g.,][]{Klochkova:1997}.  We therefore use a \citet{Castelli:2004} ATLAS9 model stellar atmosphere with $T_{\star}$ = 8250 K.  We find that to reproduce the broad, flat spectrum observed in the $\sim$ 2 $-$ 7 $\micron$ range we require the dust temperature at the inner radius of the shell be no hotter than 800 K; a hotter dust condensation temperature cannot account for this portion of the SED.  We therefore adopt 800 K and  again assume the dust to be the circumstellar silicates from \citet{Ossenkopf:1992}.  We do not attempt to reproduce the exact shape of IRC +10420's wide, flat-topped 10 $\micron$ silicate feature seen in the ISO SWS spectrum.  A flattened profile can indicate large grains with diameters approaching the micron size range \citep[e.g.,][]{van-Boekel:2005}, or self-absorption due to the optical depth approaching unity at this wavelength.   IRC +10420's complex ejecta may combine regions of silicate emission and (self) absorption which are not captured by the assumption of spherical symmetry.

\begin{figure*}
\centering
\subfloat[][]{\label{IRC_M119profile} \includegraphics[scale=0.3]{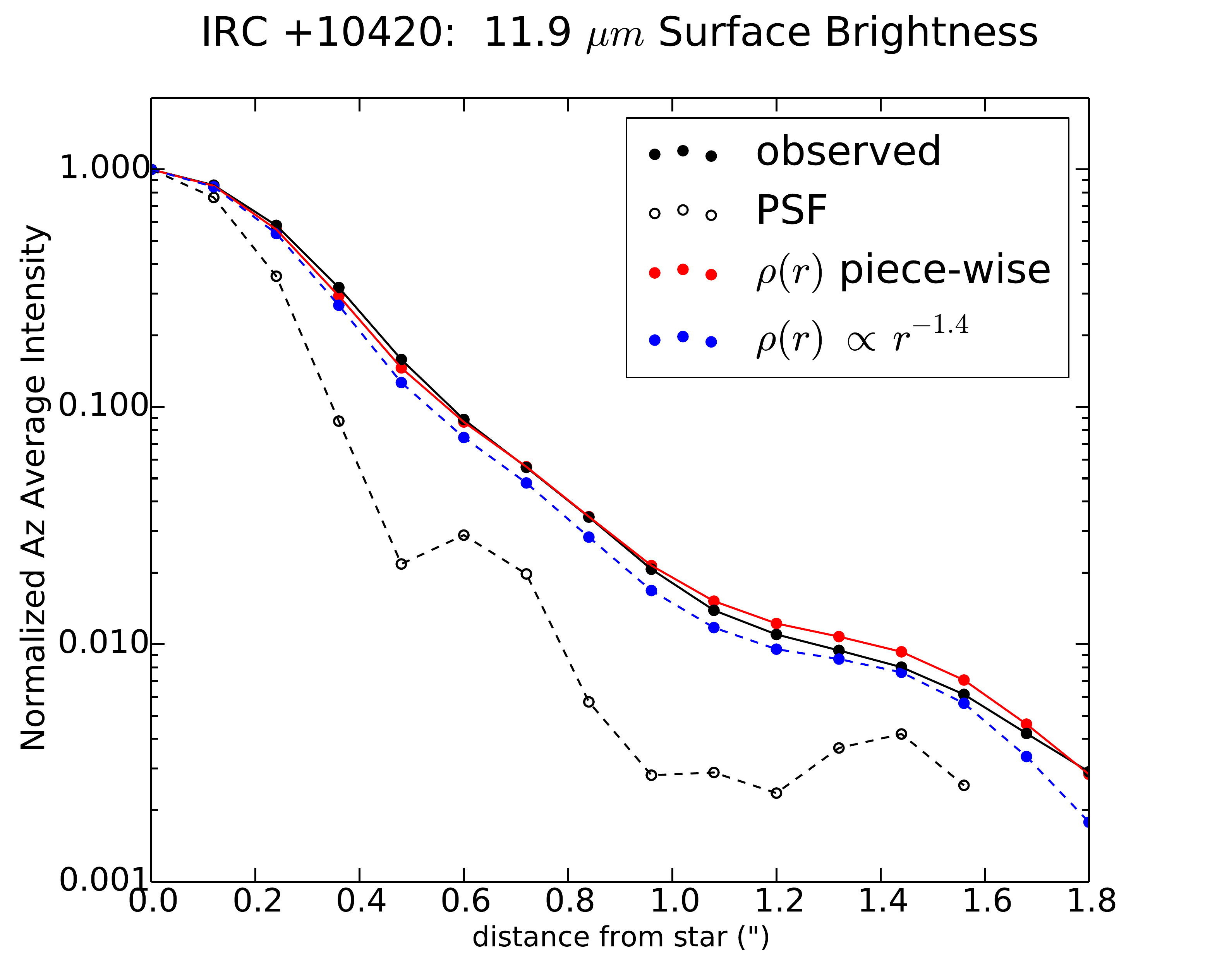}}
\subfloat[][]{\label{IRC_density_dist} \includegraphics[scale=0.33]{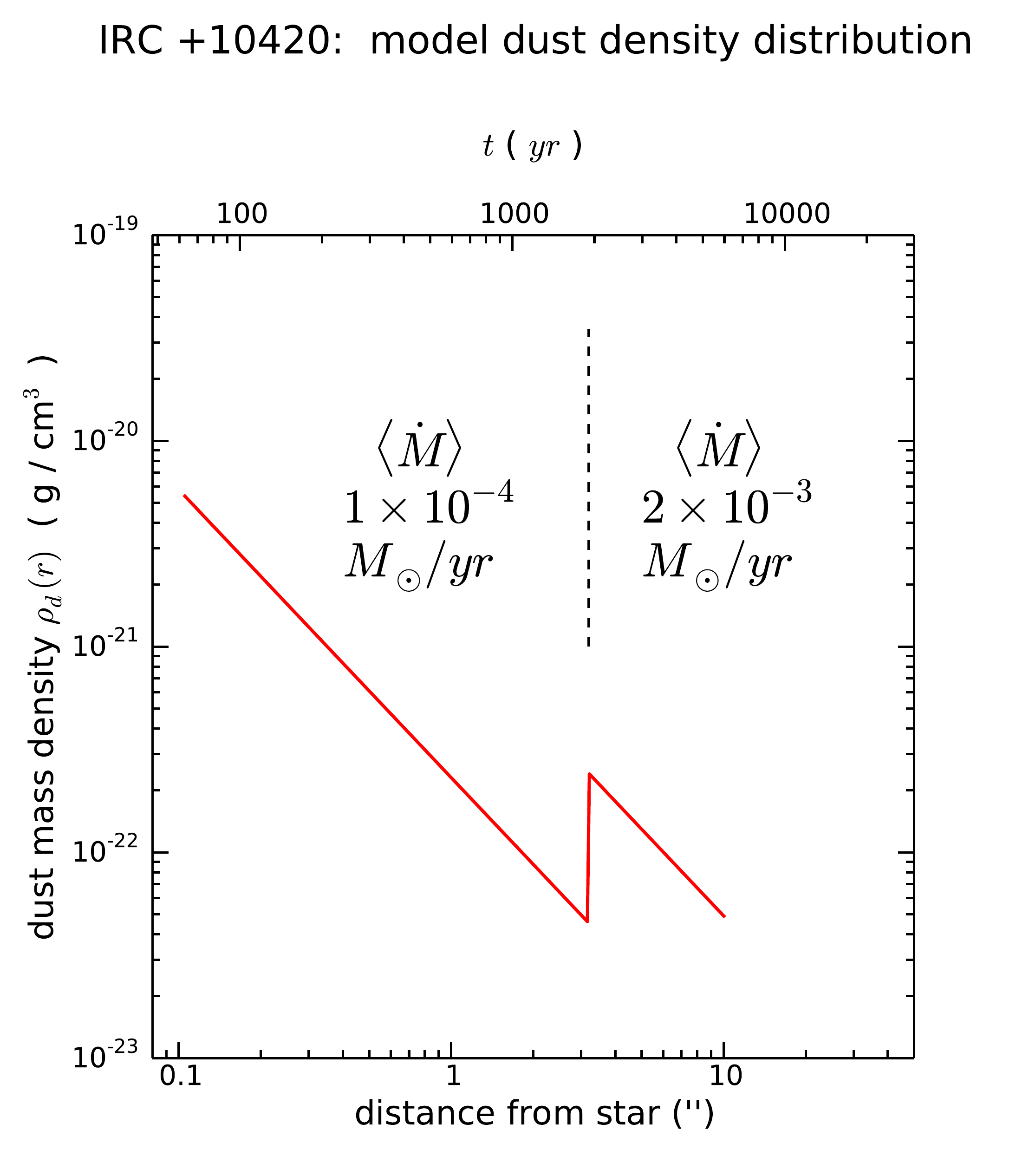}}
\caption{IRC +10420.  \textbf{(a):}  DUSTY model intensity profiles after convolution with PSF (dashed line).  The solid black line is the observed azimuthal average intensity in the MIRAC4 11.9 $\micron$ filter.  A constant mass-loss rate (not pictured) fails to produce sufficient intensity in the inner 2$\arcsec$ of IRC +10420's circumstellar environment.  A density power law index of $q$ $=$ 1.4 provides a good fit to the observed profile in this inner region.  \textbf{(b):}  The best fit DUSTY model density profile $\rho(r)$ $\propto$ $r^{-1.4}$, for which a factor of 5 enhancement in density is added at a radius of 3$\arcsec$ to account for the long-wavelength flux on the SED.  The average mass-loss rate for the earlier period spanned by the shell from 6000 to 2000 yr ago is $\langle\dot{M}\rangle$ $=$ 2 $\times$ 10$^{-3}$ $M_{\sun}$ yr$^{-1}$, while the average rate more recently is 1 $\times$ 10$^{-4}$ $M_{\sun}$ yr$^{-1}$.\\ \\} 
\end{figure*}

We first attempt to fit IRC +10420 with a constant mass-loss rate density distribution $\rho(r)$ $\propto$ $r^{-2}$ for a shell extending out to $\sim$10$\arcsec$.  Such a model is unable to account for the observed radial intensity profile out to $\sim$2$\arcsec$ at 11.9 $\micron$, in addition to failing to fit the SED beyond 12 $\micron$. Previous studies applying radiative transfer models to IRC +10420 \citep{Oudmaijer:1996, Blocker:1999, Lipman:2000} have similarly considered and rejected a constant mass-loss rate.  We find that to fit the inner 2$\arcsec$ profile at 11.9 $\micron$ requires a shallower slope to the density distribution.  We obtain a good fit to this profile with a single density power law index of $q$ $=$ 1.4 throughout the shell.  However, even with the shallower slope this still fails to produce sufficient flux in the FORCAST range and beyond.  The total energy re-radiated in that range indicates there is a substantial amount of dust with temperatures of approximately 100 $-$ 150 K.  To account for this cooler component to our model, we introduce a factor of 5 enhancement in the dust density around the star at a radius of $\sim$3$\arcsec$.  This piece-wise density profile is depicted in Figure \ref{IRC_density_dist} with a red solid line.   The actual distribution of cooler dust around the star is likely more complex than just a single second shell at this radius.  We choose to use a single enhancement as the simplest way to explain the SED in the FORCAST range rather than attempting to introduce multiple shells.  The resulting fits to the SED and 11.9 $\micron$ intensity profile are depicted with red solid lines in Figures \ref{IRC_SED} and \ref{IRC_M119profile} respectively.  

The duration of the mass-loss period attributed to the formation of the ejecta for this best-fit model depends on the expansion velocity adopted.   As with VY CMa, assuming a single expansion velocity is a simplification in the case of IRC +10420 since the expansion velocity varies throughout its envelope.  In particular, \emph{HST}/STIS long-slit spectroscopy by \citet{Humphreys:2002} demonstrated the H gas within 2$\arcsec$ is expanding at 50 $-$ 60 km s$^{-1}$ in a roughly spherical arrangement, while other spectral lines indicated a combination of outflow and infall of material at different velocities.  With this caveat in mind, we adopt as an average expansion velocity the value of 40 km s$^{-1}$ determined from CO line widths \citep{Oudmaijer:1996,De-Beck:2010}, which yields an age $\Delta t$ $=$ 6000 yr for the 10$\arcsec$ radius model shell.  We integrate each of the two parts of the density profile in Figure \ref{IRC_density_dist} with our assumed 100:1 gas-to-dust mass ratio, finding masses of 7.1 and 0.2 $M_{\sun}$ respectively in the outer and inner parts.  The resulting average mass-loss rates are $\langle\dot{M}\rangle$ $\approx$ 2 $\times$ 10$^{-3}$ $M_{\sun}$ yr$^{-1}$ for the period from 6000 until about 2000 yr ago, with a lower rate of $\sim$ 1 $\times$ 10$^{-4}$ $M_{\sun}$ yr$^{-1}$ within the past 2000 yr.  These rates are indicated on Figure \ref{IRC_density_dist} for the two periods, with time plotted on the top axis.  Exactly when the transition from a higher to a lower rate took place is admittedly somewhat uncertain; our introduction of a single, steep density enhancement 2000 yr ago is only an approximation to what was likely a more gradual transition.  It is clear however that some substantial change in the rate of mass loss has occurred in the past several thousand years, since a single shell density profile that adequately accounts for the inner 2$\arcsec$ cannot explain the large excess flux in the FORCAST range (see blue-dashed line in Figure \ref{IRC_SED}).  This supports the view that IRC +10420 is a post-RSG hypergiant which shed considerable mass during its previous stage as an RSG and will continue to evolve rapidly blueward across the upper HR Diagram.  For comparison, we note that our higher mass-loss rate for the outer ejecta is within a factor of 2 of the rate of  $3.6$ $\times$ 10$^{-3}$ $M_{\sun}$ yr$^{-1}$ found by \citet{De-Beck:2010} from modeling of multiple CO lines.

\subsection{$\rho$ Cas}
The yellow hypergiant $\rho$ Cas is famous for its historical and recent ``shell'' episodes in 1946, 1985, and 2000 during which it temporarily develops TiO bands in a cool, optically thick wind, after which it returns to its F supergiant spectrum \citep{Popper:1947, Bidelman:1957}.   During the 2000 event it shed mass at a high rate of $3\times 10^{-2} ~M_{\sun}$ for $\sim$ 200 days \citep{Lobel:2003}. The 1946 event was accompanied by a dimming in photographic magnitude of $\Delta m \approx 1.5$ for several months \citep{Beardsley:1961}, believed to be associated with a burst of ejected material.  With its spectroscopic and photometric variability, and enhanced abundances, $\rho$ Cas is considered to be a post-RSG like IRC +10420.  However, it is not a known strong infrared or maser source and $HST$ images also did not reveal any extended ejecta \citep{Schuster:2006}.   Near-IR ground-based photometry from 2.2 $-$ 12.6 $\micron$ obtained between 1969 $-$ 1973 did not show any evidence of infrared excess above the photospheric level \citep{Gillett:1970, Hackwell:1974}.  Subsequently, \citet{Jura:1990} analyzed IRAS observations taken in 1983 and found a measurable infrared excess above the photospheric level in the 12, 25 and 60 $\micron$ bands.  Even after color-correction to account for the broad bandpasses of the IRAS filters the excess flux was twice that of the photospheric level at 12 and 25 $\micron$.  They modeled this IR excess as flux from an optically thin shell of dust, presumed to have condensed between 1973 and 1983 as a result of the 1946 ejection.  Their fit to the IRAS 12, 25 and 60 $\micron$ fluxes indicated a dust temperature of 600 K for an assumed emissivity $\propto$ $\lambda^{-0.5}$, corresponding to a shell radius of 270 AU.  

\begin{figure*}
\begin{center}
\includegraphics[scale=0.38]{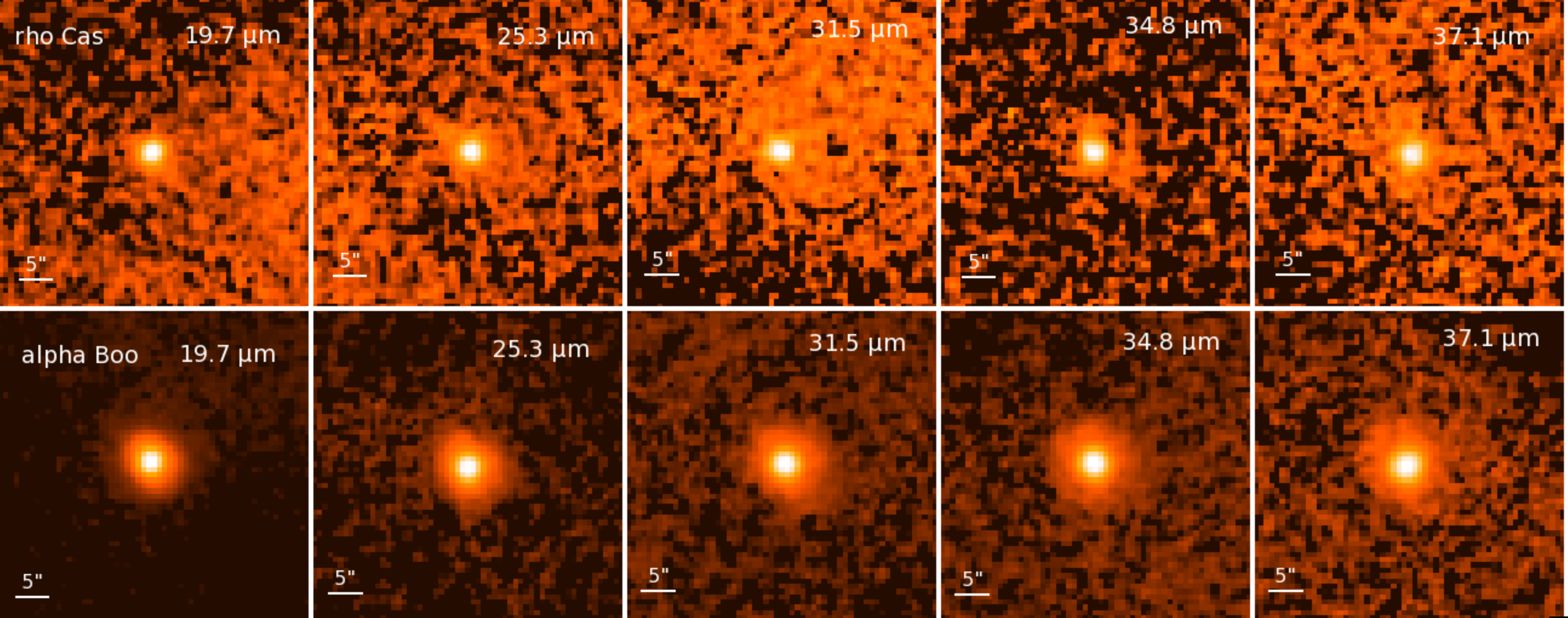}
\caption{$\rho$ Cas:  Far-IR SOFIA/FORCAST images. \textbf{Top row:}  $\rho$ Cas in each of the FORCAST filters listed in Table 1, with central wavelengths indicated in each panel.  Each f.o.v. is $45\arcsec \times 45\arcsec$ with North up, East left.  The intensity is square-root scaled.  \textbf{Bottom row:}  Standard star $\alpha$ Boo observed on the same flight (\#156).}
\label{rhoCas_ims}
\end{center}
\end{figure*}

\begin{figure*}
\begin{center}
\includegraphics[scale=0.45]{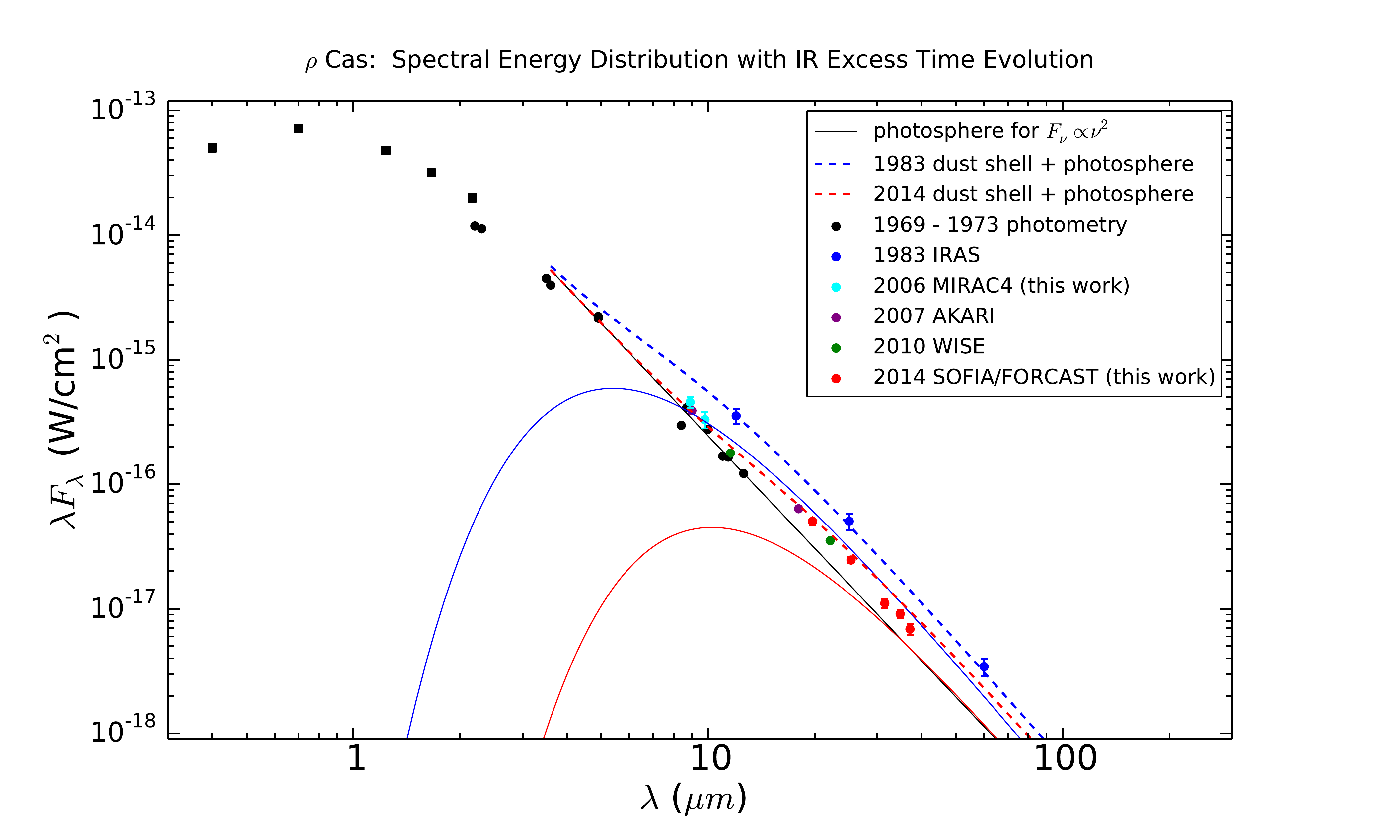}
\caption{$\rho$ Cas:  SED with time-evolution of the optically thin dust shell formed several decades after the 1946 eruption.  The black squares are photometry from the US Naval Observatory and 2MASS catalogs.  The black circles are 2 $-$12 $\micron$ ground-based photometry from \citet{Gillett:1970} and \citet{Hackwell:1974}.  The thin black solid line is the pre-IRAS photosphere level, assumed to scale as $F_{\nu}\propto \nu^{2}$ into the far-IR.  The blue circles are the color-corrected 1983 IRAS fluxes from \citet{Jura:1990}.  The thin blue solid line is the SED of the dust shell for those authors' best-fit model of dust with emissivity $\propto$ $\nu^{0.5}$ and temperature $T_{d}$ = 600 K in 1983.  The blue dashed line is the sum of the photosphere and the shell's 1983 SED.  The thin red solid line is the shell's predicted SED in 2014 per Equation (\ref{eq:shell_evol}), and the red-dashed line is the sum for the same photosphere.  The red points are the 19.7 $-$ 37.1 $\micron$ fluxes from the 2014 SOFIA/FORCAST observations.  Along with photometry from our MIRAC4 images (2006), AKARI (2007) and WISE (2010), these are generally consistent with the continuing thinning (and cooling) of the shell, with no indication of new dust formation from more recent eruptions.  The 1-$\sigma$ uncertainties are smaller than the plotted symbols sizes except for the longest FORCAST wavelengths.}
\label{rhoCas_SED}
\end{center}
\end{figure*}

We examine the time evolution of this shell model in comparison with photometry from our FORCAST images (Figure \ref{rhoCas_ims}), evolving the shell to predict its flux in 2014.  The fraction of the star's luminosity absorbed and re-radiated by the shell (the absorption optical depth) decreases as 1 / $r^{2}$ as it expands, since the solid angle subtended by each grain at the star decreases by this factor.  The shell's bolometric flux in 2014 is therefore determined by scaling its flux in 1983: 
\begin{equation}
\label{eq:shell_evol}
F_{bol,2014} = F_{bol,1983} \left( \frac{r_{1983}}{r_{2014}} \right)^{2}
\end{equation}
where radius $r_{2014}$ $=$ $r_{1983} + v_{exp}$ $\times$ (31 yr).   Jura \& Kleinmann estimated an upper limit expansion velocity of $v_{exp}$ $\approx$ 110 km s$^{-1}$, for which the radius $r_{1983}$ $=$ 270 AU would have increased to approximately 1000 AU in 2014.  $F_{bol,2014}$ would therefore be $\sim$ 7\% of $F_{bol,1983}$.  The peak of the shell's $\lambda F_{\lambda}$ spectrum would have moved to about 10 $\micron$ due to cooling of the grains to $\sim$ 310 K, as their distance from the star increased.  On the spectral energy distribution of $\rho$ Cas in Figure \ref{rhoCas_SED} the blue-dashed line is the sum of the shell's 1983 SED (thin solid blue line) and pre-IRAS photosphere (thin black line).  The red dashed-line line is the sum of the same photosphere and the predicted flux of the dust shell in 2014 (thin solid red line).  The 2014 observed FORCAST 19.7 $-$ 37.1 $\micron$ fluxes are plotted in Figure \ref{rhoCas_SED} (red points).  We also plot additional photometry taken within several years prior to the FORCAST observations:  8.9 and 9.8 $\micron$ fluxes from our MIRAC4 images in 2006; color-corrected 9 and 18 $\micron$ fluxes recorded by the AKARI satellite survey in 2007; and color-corrected fluxes at 12 and 22 $\micron$ fluxes recorded by WISE in 2010.  The predicted total flux in 2014 is generally consistent with the observed fluxes through $\sim$25 $\micron$,  though it is slightly higher than the observed flux from 25 to 40 $\micron$.  Overall the new photometry is consistent with emission from the expanding dust shell ejected in its 1946 eruption, with no evidence of newer dust formation from its more recent eruptions. Future follow-up observations would be worthwhile to see if any new excess is found in the coming decades.

\section{Summary \& Conclusions}
\underline{\emph{$\mu$ Cep}}:  We present adaptive optics (AO) diffraction-limited 8 $-$ 12 $\micron$ MMT/MIRAC3 images which reveal a circumstellar envelope extending approximately East-West out to $\sim$ 0$\farcs$5 (several tens of stellar radii).  This envelope is the likely source of the  10 $\micron$ silicate feature in $\mu$ Cep's SED.  Our SOFIA/FORCAST far-IR imaging of $\mu$ Cep shows extended emission from 25 $-$ 37 $\micron$ out to $\sim20\arcsec$ around the star, with further extension to $\sim25\arcsec$ to the East in the same direction as the extensive nebula previously revealed in \emph{Herschel} PACS 70 \& 160 $\micron$ images \citep{Cox:2012}.  

DUSTY radiative transfer modeling of the 37.1 $\micron$ profile and the SED indicate an average mass-loss rate of 4 $\times$ 10$^{-6}$ $M_{\sun}$ yr $^{-1}$, with evidence for a slow decline in mass-loss rate over the 13,000 yr dynamical age of its shell.  The range of mass-loss rates is consistent with previous studies finding $\mu$ Cep's rate is significantly lower than RSGs of comparable luminosity.

\underline{\emph{VY CMa}}:   SOFIA/FORCAST images from 19.7 $-$ 37.1 $\micron$ display a morphology which coincides with the general shape of the highly asymmetric nebulae seen in the visual, suggesting thermal emission from dust associated with the expanding arcs to the northwest and southwest.  Our best-fit DUSTY model indicates an average mass-loss rate of 6 $\times$ 10$^{-4}$ $M_{\sun}$ yr$^{-1}$.

\underline{\emph{IRC +10420}}:  We present adaptive optics (AO) diffraction-limited 8 $-$ 12 $\micron$ MMT/MIRAC4 images which reveal spatially extended, circularly symmetric emission out to a radius of $\sim2\arcsec$.  As with VY CMa, the extended shape of IRC +10420 in our SOFIA/FORCAST images coincides with the general shape of the asymmetric nebulae seen in the visual.  Our best-fit DUSTY model indicates a change in the average mass-loss rate, with a high average rate of 2 $\times$ 10$^{-3}$ $M_{\sun}$ yr$^{-1}$ from 6000 $-$ 2000 yr ago during its presumed RSG stage, and a lower average rate of 1 $\times$ 10$^{-4}$ $M_{\sun}$ yr$^{-1}$ in the past 2000 yr.

\underline{\emph{$\rho$ Cas}}:  Our new 19.7 $-$ 37.1 $\micron$ SOFIA/FORCAST photometry of this yellow hypergiant are consistent with the continued expansion and thinning of the dust shell fitted to the infrared excess observed with IRAS, which was attributed to dust formed as a result of the 1946 eruption.  There is no indication of new dust formation from more recent eruptions.

\section{Acknowledgements}
We thank Dr. Willem-Jan de Wit, Dr. Takuya Fujiyoshi and the Subaru/COMICS instrument team for consulting on the orientation of $\mu$ Cep's nebula as observed at 24.5 $\micron$.  This work has used unpublished data from Michael Schuster's PhD thesis, which is available through the SAO/NASA Astrophysics Data System (ADS) at http://adsabs.harvard.edu/abs/2007PhDT........28S.  Financial support for this work was provided by NASA through award \# SOF-0091 to R. M. Humphreys issued by USRA.

\bibliographystyle{apj}
\bibliography{ms_preprint.bbl}

\begin{thebibliography}{}
\expandafter\ifx\csname natexlab\endcsname\relax\def\natexlab#1{#1}\fi

\bibitem[{{Beardsley}(1961)}]{Beardsley:1961}
{Beardsley}, W.~R. 1961, \apjs, 5, 381

\bibitem[{{Bidelman} \& {McKellar}(1957)}]{Bidelman:1957}
{Bidelman}, W.~P., \& {McKellar}, A. 1957, \pasp, 69, 31

\bibitem[{{Biller} {et~al.}(2005){Biller}, {Close}, {Li}, {Bieging},
  {Hoffmann}, {Hinz}, {Miller}, {Brusa}, {Lloyd-Hart}, {Wildi}, {Potter}, \&
  {Oppenheimer}}]{Biller:2005}
{Biller}, B.~A., {Close}, L.~M., {Li}, A., {et~al.} 2005, \apj, 620, 450

\bibitem[{{Bl{\"o}cker} {et~al.}(1999){Bl{\"o}cker}, {Balega}, {Hofmann},
  {Lichtenth{\"a}ler}, {Osterbart}, \& {Weigelt}}]{Blocker:1999}
{Bl{\"o}cker}, T., {Balega}, Y., {Hofmann}, K.-H., {et~al.} 1999, \aap, 348,
  805

\bibitem[{{Castelli} \& {Kurucz}(2004)}]{Castelli:2004}
{Castelli}, F., \& {Kurucz}, R.~L. 2004, ArXiv Astrophysics e-prints,
  astro-ph/0405087

\bibitem[{{Clegg} {et~al.}(1996){Clegg}, {Ade}, {Armand}, {Baluteau}, {Barlow},
  {Buckley}, {Berges}, {Burgdorf}, {Caux}, {Ceccarelli}, {Cerulli}, {Church},
  {Cotin}, {Cox}, {Cruvellier}, {Culhane}, {Davis}, {di Giorgio}, {Diplock},
  {Drummond}, {Emery}, {Ewart}, {Fischer}, {Furniss}, {Glencross},
  {Greenhouse}, {Griffin}, {Gry}, {Harwood}, {Hazell}, {Joubert}, {King},
  {Lim}, {Liseau}, {Long}, {Lorenzetti}, {Molinari}, {Murray}, {Naylor},
  {Nisini}, {Norman}, {Omont}, {Orfei}, {Patrick}, {Pequignot}, {Pouliquen},
  {Price}, {Nguyen-Q-Rieu}, {Rogers}, {Robinson}, {Saisse}, {Saraceno},
  {Serra}, {Sidher}, {Smith}, {Smith}, {Spinoglio}, {Swinyard}, {Texier},
  {Towlson}, {Trams}, {Unger}, \& {White}}]{Clegg:1996}
{Clegg}, P.~E., {Ade}, P.~A.~R., {Armand}, C., {et~al.} 1996, \aap, 315, L38

\bibitem[{{Cox} {et~al.}(2012){Cox}, {Kerschbaum}, {van Marle}, {Decin},
  {Ladjal}, {Mayer}, {Groenewegen}, {van Eck}, {Royer}, {Ottensamer}, {Ueta},
  {Jorissen}, {Mecina}, {Meliani}, {Luntzer}, {Blommaert}, {Posch},
  {Vandenbussche}, \& {Waelkens}}]{Cox:2012}
{Cox}, N.~L.~J., {Kerschbaum}, F., {van Marle}, A.-J., {et~al.} 2012, \aap,
  537, A35

\bibitem[{{Danchi} {et~al.}(1994){Danchi}, {Bester}, {Degiacomi}, {Greenhill},
  \& {Townes}}]{Danchi:1994}
{Danchi}, W.~C., {Bester}, M., {Degiacomi}, C.~G., {Greenhill}, L.~J., \&
  {Townes}, C.~H. 1994, \aj, 107, 1469

\bibitem[{{De Beck} {et~al.}(2010){De Beck}, {Decin}, {de Koter}, {Justtanont},
  {Verhoelst}, {Kemper}, \& {Menten}}]{De-Beck:2010}
{De Beck}, E., {Decin}, L., {de Koter}, A., {et~al.} 2010, \aap, 523, A18

\bibitem[{{De Beck} {et~al.}(2015){De Beck}, {Vlemmings}, {Muller}, {Black},
  {O'Gorman}, {Richards}, {Baudry}, {Maercker}, {Decin}, \&
  {Humphreys}}]{De-Beck:2015}
{De Beck}, E., {Vlemmings}, W., {Muller}, S., {et~al.} 2015, \aap, 580, A36

\bibitem[{{de Graauw} {et~al.}(1996){de Graauw}, {Haser}, {Beintema},
  {Roelfsema}, {van Agthoven}, {Barl}, {Bauer}, {Bekenkamp}, {Boonstra},
  {Boxhoorn}, {Cote}, {de Groene}, {van Dijkhuizen}, {Drapatz}, {Evers},
  {Feuchtgruber}, {Frericks}, {Genzel}, {Haerendel}, {Heras}, {van der Hucht},
  {van der Hulst}, {Huygen}, {Jacobs}, {Jakob}, {Kamperman}, {Katterloher},
  {Kester}, {Kunze}, {Kussendrager}, {Lahuis}, {Lamers}, {Leech}, {van der
  Lei}, {van der Linden}, {Luinge}, {Lutz}, {Melzner}, {Morris}, {van Nguyen},
  {Ploeger}, {Price}, {Salama}, {Schaeidt}, {Sijm}, {Smoorenburg}, {Spakman},
  {Spoon}, {Steinmayer}, {Stoecker}, {Valentijn}, {Vandenbussche}, {Visser},
  {Waelkens}, {Waters}, {Wensink}, {Wesselius}, {Wiezorrek}, {Wieprecht},
  {Wijnbergen}, {Wildeman}, \& {Young}}]{de-Graauw:1996}
{de Graauw}, T., {Haser}, L.~N., {Beintema}, D.~A., {et~al.} 1996, \aap, 315,
  L49

\bibitem[{{de Wit} {et~al.}(2008){de Wit}, {Oudmaijer}, {Fujiyoshi}, {Hoare},
  {Honda}, {Kataza}, {Miyata}, {Okamoto}, {Onaka}, {Sako}, \&
  {Yamashita}}]{de-Wit:2008}
{de Wit}, W.~J., {Oudmaijer}, R.~D., {Fujiyoshi}, T., {et~al.} 2008, \apjl,
  685, L75

\bibitem[{{Dinh-V.-Trung} {et~al.}(2009){Dinh-V.-Trung}, {Muller}, {Lim},
  {Kwok}, \& {Muthu}}]{Dinh-V.-Trung:2009}
{Dinh-V.-Trung}, {Muller}, S., {Lim}, J., {Kwok}, S., \& {Muthu}, C. 2009,
  \apj, 697, 409

\bibitem[{{Gehrz} \& {Woolf}(1971)}]{Gehrz:1971}
{Gehrz}, R.~D., \& {Woolf}, N.~J. 1971, \apj, 165, 285

\bibitem[{{Gillett} {et~al.}(1970){Gillett}, {Hyland}, \&
  {Stein}}]{Gillett:1970}
{Gillett}, F.~C., {Hyland}, A.~R., \& {Stein}, W.~A. 1970, \apjl, 162, L21

\bibitem[{{Groenewegen} {et~al.}(2011){Groenewegen}, {Waelkens}, {Barlow},
  {Kerschbaum}, {Garcia-Lario}, {Cernicharo}, {Blommaert}, {Bouwman}, {Cohen},
  {Cox}, {Decin}, {Exter}, {Gear}, {Gomez}, {Hargrave}, {Henning},
  {Hutsem{\'e}kers}, {Ivison}, {Jorissen}, {Krause}, {Ladjal}, {Leeks}, {Lim},
  {Matsuura}, {Naz{\'e}}, {Olofsson}, {Ottensamer}, {Polehampton}, {Posch},
  {Rauw}, {Royer}, {Sibthorpe}, {Swinyard}, {Ueta}, {Vamvatira-Nakou},
  {Vandenbussche}, {van de Steene}, {van Eck}, {van Hoof}, {van Winckel},
  {Verdugo}, \& {Wesson}}]{Groenewegen:2011}
{Groenewegen}, M.~A.~T., {Waelkens}, C., {Barlow}, M.~J., {et~al.} 2011, \aap,
  526, A162

\bibitem[{{Hackwell} \& {Gehrz}(1974)}]{Hackwell:1974}
{Hackwell}, J.~A., \& {Gehrz}, R.~D. 1974, \apj, 194, 49

\bibitem[{{Harwit} {et~al.}(2001){Harwit}, {Malfait}, {Decin}, {Waelkens},
  {Feuchtgruber}, \& {Melnick}}]{Harwit:2001}
{Harwit}, M., {Malfait}, K., {Decin}, L., {et~al.} 2001, \apj, 557, 844

\bibitem[{{Herter} {et~al.}(2012){Herter}, {Adams}, {De Buizer}, {Gull},
  {Schoenwald}, {Henderson}, {Keller}, {Nikola}, {Stacey}, \&
  {Vacca}}]{Herter:2012}
{Herter}, T.~L., {Adams}, J.~D., {De Buizer}, J.~M., {et~al.} 2012, \apjl, 749,
  L18

\bibitem[{{Herter} {et~al.}(2013){Herter}, {Vacca}, {Adams}, {Keller},
  {Schoenwald}, {Hirsch}, {Wang}, {De Buizer}, {Helton}, \&
  {Llorens}}]{Herter:2013}
{Herter}, T.~L., {Vacca}, W.~D., {Adams}, J.~D., {et~al.} 2013, \pasp, 125,
  1393

\bibitem[{{Heske}(1990)}]{Heske:1990}
{Heske}, A. 1990, \aap, 229, 494

\bibitem[{{Hinz} {et~al.}(2000){Hinz}, {Angel}, {Woolf}, {Hoffmann}, \&
  {McCarthy}}]{Hinz:2000}
{Hinz}, P.~M., {Angel}, J.~R.~P., {Woolf}, N.~J., {Hoffmann}, W.~F., \&
  {McCarthy}, D.~W. 2000, in Society of Photo-Optical Instrumentation Engineers
  (SPIE) Conference Series, Vol. 4006, Interferometry in Optical Astronomy, ed.
  P.~{L{\'e}na} \& A.~{Quirrenbach}, 349--353

\bibitem[{{Hoffmann} {et~al.}(1998){Hoffmann}, {Hora}, {Fazio}, {Deutsch}, \&
  {Dayal}}]{Hoffmann:1998}
{Hoffmann}, W.~F., {Hora}, J.~L., {Fazio}, G.~G., {Deutsch}, L.~K., \& {Dayal},
  A. 1998, in Society of Photo-Optical Instrumentation Engineers (SPIE)
  Conference Series, Vol. 3354, Infrared Astronomical Instrumentation, ed.
  A.~M. {Fowler}, 647--658

\bibitem[{{Humphreys}(1978)}]{Humphreys:1978}
{Humphreys}, R.~M. 1978, \apjs, 38, 309

\bibitem[{{Humphreys} {et~al.}(2005){Humphreys}, {Davidson}, {Ruch}, \&
  {Wallerstein}}]{Humphreys:2005}
{Humphreys}, R.~M., {Davidson}, K., {Ruch}, G., \& {Wallerstein}, G. 2005, \aj,
  129, 492

\bibitem[{{Humphreys} {et~al.}(2002){Humphreys}, {Davidson}, \&
  {Smith}}]{Humphreys:2002}
{Humphreys}, R.~M., {Davidson}, K., \& {Smith}, N. 2002, \aj, 124, 1026

\bibitem[{{Humphreys} {et~al.}(2007){Humphreys}, {Helton}, \&
  {Jones}}]{Humphreys:2007}
{Humphreys}, R.~M., {Helton}, L.~A., \& {Jones}, T.~J. 2007, \aj, 133, 2716

\bibitem[{{Humphreys} {et~al.}(1997){Humphreys}, {Smith}, {Davidson}, {Jones},
  {Gehrz}, {Mason}, {Hayward}, {Houck}, \& {Krautter}}]{Humphreys:1997}
{Humphreys}, R.~M., {Smith}, N., {Davidson}, K., {et~al.} 1997, \aj, 114, 2778

\bibitem[{{Ivezic} \& {Elitzur}(1997)}]{Ivezic:1997}
{Ivezic}, Z., \& {Elitzur}, M. 1997, \mnras, 287, 799

\bibitem[{{Jones} {et~al.}(2007){Jones}, {Humphreys}, {Helton}, {Gui}, \&
  {Huang}}]{Jones:2007}
{Jones}, T.~J., {Humphreys}, R.~M., {Helton}, L.~A., {Gui}, C., \& {Huang}, X.
  2007, \aj, 133, 2730

\bibitem[{{Jones} {et~al.}(1993){Jones}, {Humphreys}, {Gehrz}, {Lawrence},
  {Zickgraf}, {Moseley}, {Casey}, {Glaccum}, {Koch}, {Pina}, {Jones}, {Venn},
  {Stahl}, \& {Starrfield}}]{Jones:1993}
{Jones}, T.~J., {Humphreys}, R.~M., {Gehrz}, R.~D., {et~al.} 1993, \apj, 411,
  323

\bibitem[{{Jura} \& {Kleinmann}(1990{\natexlab{a}})}]{Jura:1990a}
{Jura}, M., \& {Kleinmann}, S.~G. 1990{\natexlab{a}}, \apjs, 73, 769

\bibitem[{{Jura} \& {Kleinmann}(1990{\natexlab{b}})}]{Jura:1990}
---. 1990{\natexlab{b}}, \apj, 351, 583

\bibitem[{{Justtanont} {et~al.}(1997){Justtanont}, {Yamamura}, {de Jong}, \&
  {Waters}}]{Justtanont:1997}
{Justtanont}, K., {Yamamura}, I., {de Jong}, T., \& {Waters}, L.~B.~F.~M. 1997,
  \apss, 251, 25

\bibitem[{{Kastner} \& {Weintraub}(1995)}]{Kastner:1995}
{Kastner}, J.~H., \& {Weintraub}, D.~A. 1995, \apj, 452, 833

\bibitem[{{Klochkova} {et~al.}(1997){Klochkova}, {Chentsov}, \&
  {Panchuk}}]{Klochkova:1997}
{Klochkova}, V.~G., {Chentsov}, E.~L., \& {Panchuk}, V.~E. 1997, \mnras, 292,
  19

\bibitem[{{Knapp}(1985)}]{Knapp:1985a}
{Knapp}, G.~R. 1985, \apj, 293, 273

\bibitem[{{Knapp} {et~al.}(1993){Knapp}, {Sandell}, \& {Robson}}]{Knapp:1993}
{Knapp}, G.~R., {Sandell}, G., \& {Robson}, E.~I. 1993, \apjs, 88, 173

\bibitem[{{Ladjal} {et~al.}(2010){Ladjal}, {Justtanont}, {Groenewegen},
  {Blommaert}, {Waelkens}, \& {Barlow}}]{Ladjal:2010}
{Ladjal}, D., {Justtanont}, K., {Groenewegen}, M.~A.~T., {et~al.} 2010, \aap,
  513, A53

\bibitem[{{Lee}(1970)}]{Lee:1970}
{Lee}, T.~A. 1970, \pasp, 82, 765

\bibitem[{{Levesque} {et~al.}(2005){Levesque}, {Massey}, {Olsen}, {Plez},
  {Josselin}, {Maeder}, \& {Meynet}}]{Levesque:2005}
{Levesque}, E.~M., {Massey}, P., {Olsen}, K.~A.~G., {et~al.} 2005, \apj, 628,
  973

\bibitem[{{Lipman} {et~al.}(2000){Lipman}, {Hale}, {Monnier}, {Tuthill},
  {Danchi}, \& {Townes}}]{Lipman:2000}
{Lipman}, E.~A., {Hale}, D.~D.~S., {Monnier}, J.~D., {et~al.} 2000, \apj, 532,
  467

\bibitem[{{Lloyd} {et~al.}(2003){Lloyd}, {Lerate}, \& {Grundy}}]{Lloyd:2003}
{Lloyd}, C., {Lerate}, M.~R., \& {Grundy}, T.~W. 2003, {ISO Technical Note 17
  (Madrid: ISO Data Centre)}

\bibitem[{{Lobel} {et~al.}(2003){Lobel}, {Dupree}, {Stefanik}, {Torres},
  {Israelian}, {Morrison}, {de Jager}, {Nieuwenhuijzen}, {Ilyin}, \&
  {Musaev}}]{Lobel:2003}
{Lobel}, A., {Dupree}, A.~K., {Stefanik}, R.~P., {et~al.} 2003, \apj, 583, 923

\bibitem[{{Mathis} {et~al.}(1977){Mathis}, {Rumpl}, \&
  {Nordsieck}}]{Mathis:1977}
{Mathis}, J.~S., {Rumpl}, W., \& {Nordsieck}, K.~H. 1977, \apj, 217, 425

\bibitem[{{Mauron} \& {Josselin}(2011)}]{Mauron:2011}
{Mauron}, N., \& {Josselin}, E. 2011, \aap, 526, A156

\bibitem[{{Meixner} {et~al.}(1999){Meixner}, {Ueta}, {Dayal}, {Hora}, {Fazio},
  {Hrivnak}, {Skinner}, {Hoffmann}, \& {Deutsch}}]{Meixner:1999}
{Meixner}, M., {Ueta}, T., {Dayal}, A., {et~al.} 1999, \apjs, 122, 221

\bibitem[{{Montez} {et~al.}(2015){Montez}, {Kastner}, {Humphreys}, {Turok}, \&
  {Davidson}}]{Montez:2015}
{Montez}, Jr., R., {Kastner}, J.~H., {Humphreys}, R.~M., {Turok}, R.~L., \&
  {Davidson}, K. 2015, \apj, 800, 4

\bibitem[{{Neckel} {et~al.}(1980){Neckel}, {Klare}, \&
  {Sarcander}}]{Neckel:1980}
{Neckel}, T., {Klare}, G., \& {Sarcander}, M. 1980, \aaps, 42, 251

\bibitem[{{O'Gorman} {et~al.}(2015){O'Gorman}, {Vlemmings}, {Richards},
  {Baudry}, {De Beck}, {Decin}, {Harper}, {Humphreys}, {Kervella}, {Khouri}, \&
  {Muller}}]{OGorman:2015}
{O'Gorman}, E., {Vlemmings}, W., {Richards}, A.~M.~S., {et~al.} 2015, \aap,
  573, L1

\bibitem[{{Ossenkopf} {et~al.}(1992){Ossenkopf}, {Henning}, \&
  {Mathis}}]{Ossenkopf:1992}
{Ossenkopf}, V., {Henning}, T., \& {Mathis}, J.~S. 1992, \aap, 261, 567

\bibitem[{{Oudmaijer} \& {de Wit}(2013)}]{Oudmaijer:2013}
{Oudmaijer}, R.~D., \& {de Wit}, W.~J. 2013, \aap, 551, A69

\bibitem[{{Oudmaijer} {et~al.}(1996){Oudmaijer}, {Groenewegen}, {Matthews},
  {Blommaert}, \& {Sahu}}]{Oudmaijer:1996}
{Oudmaijer}, R.~D., {Groenewegen}, M.~A.~T., {Matthews}, H.~E., {Blommaert},
  J.~A.~D.~L., \& {Sahu}, K.~C. 1996, \mnras, 280, 1062

\bibitem[{{Pilbratt} {et~al.}(2010){Pilbratt}, {Riedinger}, {Passvogel},
  {Crone}, {Doyle}, {Gageur}, {Heras}, {Jewell}, {Metcalfe}, {Ott}, \&
  {Schmidt}}]{Pilbratt:2010}
{Pilbratt}, G.~L., {Riedinger}, J.~R., {Passvogel}, T., {et~al.} 2010, \aap,
  518, L1

\bibitem[{{Poglitsch} {et~al.}(2010){Poglitsch}, {Waelkens}, {Geis},
  {Feuchtgruber}, {Vandenbussche}, {Rodriguez}, {Krause}, {Renotte}, {van
  Hoof}, {Saraceno}, {Cepa}, {Kerschbaum}, {Agn{\`e}se}, {Ali}, {Altieri},
  {Andreani}, {Augueres}, {Balog}, {Barl}, {Bauer}, {Belbachir}, {Benedettini},
  {Billot}, {Boulade}, {Bischof}, {Blommaert}, {Callut}, {Cara}, {Cerulli},
  {Cesarsky}, {Contursi}, {Creten}, {De Meester}, {Doublier}, {Doumayrou},
  {Duband}, {Exter}, {Genzel}, {Gillis}, {Gr{\"o}zinger}, {Henning},
  {Herreros}, {Huygen}, {Inguscio}, {Jakob}, {Jamar}, {Jean}, {de Jong},
  {Katterloher}, {Kiss}, {Klaas}, {Lemke}, {Lutz}, {Madden}, {Marquet},
  {Martignac}, {Mazy}, {Merken}, {Montfort}, {Morbidelli}, {M{\"u}ller},
  {Nielbock}, {Okumura}, {Orfei}, {Ottensamer}, {Pezzuto}, {Popesso},
  {Putzeys}, {Regibo}, {Reveret}, {Royer}, {Sauvage}, {Schreiber}, {Stegmaier},
  {Schmitt}, {Schubert}, {Sturm}, {Thiel}, {Tofani}, {Vavrek}, {Wetzstein},
  {Wieprecht}, \& {Wiezorrek}}]{Poglitsch:2010}
{Poglitsch}, A., {Waelkens}, C., {Geis}, N., {et~al.} 2010, \aap, 518, L2

\bibitem[{{Polehampton} {et~al.}(2010){Polehampton}, {Menten}, {van der Tak},
  \& {White}}]{Polehampton:2010}
{Polehampton}, E.~T., {Menten}, K.~M., {van der Tak}, F.~F.~S., \& {White},
  G.~J. 2010, \aap, 510, A80

\bibitem[{{Popper}(1947)}]{Popper:1947}
{Popper}, D.~M. 1947, \aj, 52, 129

\bibitem[{{Richards} {et~al.}(1998){Richards}, {Yates}, \&
  {Cohen}}]{Richards:1998}
{Richards}, A.~M.~S., {Yates}, J.~A., \& {Cohen}, R.~J. 1998, \mnras, 299, 319

\bibitem[{{Rowles} \& {Froebrich}(2009)}]{Rowles:2009}
{Rowles}, J., \& {Froebrich}, D. 2009, \mnras, 395, 1640

\bibitem[{{Schuster}(2007)}]{Schuster:2007}
{Schuster}, M.~T. 2007, PhD thesis, University of Minnesota

\bibitem[{{Schuster} {et~al.}(2006){Schuster}, {Humphreys}, \&
  {Marengo}}]{Schuster:2006}
{Schuster}, M.~T., {Humphreys}, R.~M., \& {Marengo}, M. 2006, \aj, 131, 603

\bibitem[{{Schuster} {et~al.}(2009){Schuster}, {Marengo}, {Hora}, {Fazio},
  {Humphreys}, {Gehrz}, {Hinz}, {Kenworthy}, \& {Hoffmann}}]{Schuster:2009}
{Schuster}, M.~T., {Marengo}, M., {Hora}, J.~L., {et~al.} 2009, \apj, 699, 1423

\bibitem[{{Shenoy} {et~al.}(2015){Shenoy}, {Jones}, {Packham}, \&
  {Lopez-Rodriguez}}]{Shenoy:2015}
{Shenoy}, D.~P., {Jones}, T.~J., {Packham}, C., \& {Lopez-Rodriguez}, E. 2015,
  \aj, 150, 15

\bibitem[{{Shenoy} {et~al.}(2013){Shenoy}, {Jones}, {Humphreys}, {Marengo},
  {Leisenring}, {Nelson}, {Wilson}, {Skrutskie}, {Hinz}, {Hoffmann}, {Bailey},
  {Skemer}, {Rodigas}, \& {Vaitheeswaran}}]{Shenoy:2013}
{Shenoy}, D.~P., {Jones}, T.~J., {Humphreys}, R.~M., {et~al.} 2013, \aj, 146,
  90

\bibitem[{{Skemer} {et~al.}(2008){Skemer}, {Close}, {Hinz}, {Hoffmann},
  {Kenworthy}, \& {Miller}}]{Skemer:2008}
{Skemer}, A.~J., {Close}, L.~M., {Hinz}, P.~M., {et~al.} 2008, \apj, 676, 1082

\bibitem[{{Sloan} {et~al.}(2003){Sloan}, {Kraemer}, {Price}, \&
  {Shipman}}]{Sloan:2003}
{Sloan}, G.~C., {Kraemer}, K.~E., {Price}, S.~D., \& {Shipman}, R.~F. 2003,
  \apjs, 147, 379

\bibitem[{{Smith} {et~al.}(2001){Smith}, {Humphreys}, {Davidson}, {Gehrz},
  {Schuster}, \& {Krautter}}]{Smith:2001}
{Smith}, N., {Humphreys}, R.~M., {Davidson}, K., {et~al.} 2001, \aj, 121, 1111

\bibitem[{{Tiffany} {et~al.}(2010){Tiffany}, {Humphreys}, {Jones}, \&
  {Davidson}}]{Tiffany:2010}
{Tiffany}, C., {Humphreys}, R.~M., {Jones}, T.~J., \& {Davidson}, K. 2010, \aj,
  140, 339

\bibitem[{{van Boekel} {et~al.}(2005){van Boekel}, {Min}, {Waters}, {de Koter},
  {Dominik}, {van den Ancker}, \& {Bouwman}}]{van-Boekel:2005}
{van Boekel}, R., {Min}, M., {Waters}, L.~B.~F.~M., {et~al.} 2005, \aap, 437,
  189

\bibitem[{{Verhoelst} {et~al.}(2009){Verhoelst}, {van der Zypen}, {Hony},
  {Decin}, {Cami}, \& {Eriksson}}]{Verhoelst:2009}
{Verhoelst}, T., {van der Zypen}, N., {Hony}, S., {et~al.} 2009, \aap, 498, 127

\bibitem[{{Wittkowski} {et~al.}(2012){Wittkowski}, {Hauschildt},
  {Arroyo-Torres}, \& {Marcaide}}]{Wittkowski:2012}
{Wittkowski}, M., {Hauschildt}, P.~H., {Arroyo-Torres}, B., \& {Marcaide},
  J.~M. 2012, \aap, 540, L12

\bibitem[{{Zhang} {et~al.}(2012){Zhang}, {Reid}, {Menten}, \&
  {Zheng}}]{Zhang:2012}
{Zhang}, B., {Reid}, M.~J., {Menten}, K.~M., \& {Zheng}, X.~W. 2012, \apj, 744,
  23

\bibitem[{{Ziurys} {et~al.}(2007){Ziurys}, {Milam}, {Apponi}, \&
  {Woolf}}]{Ziurys:2007}
{Ziurys}, L.~M., {Milam}, S.~N., {Apponi}, A.~J., \& {Woolf}, N.~J. 2007, \nat,
  447, 1094

\end{thebibliography}

\begin{turnpage}
\begin{deluxetable*}{cccccccccccc}
\tabletypesize{\footnotesize}
\tablewidth{550pt}
\tablecaption{DUSTY Model Inputs \& Average Mass-Loss Rates}	
\tablenum{3}
\tablehead{\colhead{Target} & \colhead{$D$} & \colhead{$T_{\star}$} & \colhead{$r_{1}$} & \colhead{$r_{2}$} & \colhead{$T_{d}(r_{1})$} & \colhead{$\tau_{37.1}$} & \colhead{$\kappa_{37.1}$\tablenotemark{a}} &  \colhead{$v_{exp}$\tablenotemark{b}} & \colhead{$\Delta t$\tablenotemark{c}} &   \colhead{$\langle\dot{M}\rangle$\tablenotemark{d}}  \\ 
\colhead{} & \colhead{(kpc)} & \colhead{(K)} & \colhead{(arcsec, AU)} & \colhead{(arcsec, AU)} & \colhead{(K)} & \colhead{} & \colhead{(cm$^{2}$ g$^{-1}$)} &  \colhead{(km s$^{-1}$)} & \colhead{(yr)} & \colhead{($M_{\sun}$ yr$^{-1}$)} }
\startdata 
$\mu$ Cep    & 0.87 & 3750 & 0$\farcs$11 & 110$\arcsec$ & 1000 &  0.0029 & 194 & 35 & 13000 & (3.8 $\pm$ 0.3) $\times$ 10$^{-6}$\\
                     &          &          &     (96 AU)  & (96000 AU)    &          &               &       &       &            &    \\
                     &          &          &                    &                       &          &               &       &       &            &    \\
VY CMa      & 1.2    & 3500 &   0$\farcs$12 & 10$\arcsec$  & 800   & 0.165    & 195 &  47 & 1200 & (5.6 $\pm$ 0.6) $\times$ 10$^{-4}$ \\
                     &          &          &    (140 AU)   &  (12000 AU)  &          &               &       &        &            &     \\
                     &          &          &                    &                       &          &               &       &        &            &     \\
IRC +10420 & 5       &  8250 &   0$\farcs$10 & 10$\arcsec$& 800   & 0.026    & 194 &  40 & 2000 & (1.3 $\pm$ 0.3) $\times$ 10$^{-4}$ \\
(past 2000 yr)&         &         &    (500 AU) &  (50000 AU)    &            &              &       &        &             &  \\
                      &         &          &                    &                      &            &              &       &        &             &  \\
IRC +10420 &          &          &                    &                      &            &              &       &         & 4000 & (1.8 $\pm$ 0.3) $\times$ 10$^{-3}$ \\
($>$ 2000 yr ago) & &         &                      &                      &            &             &       &        &              & \\
\enddata
\tablecomments{Columns $r_{1}$ and $r_{2}$ are the inner and outer radii of the model shell, with $T_{d}(r_{1})$ being the assumed dust temperature at $r_{1}$.}
\tablenotetext{a}{~Opacity at 37.1 $\micron$ for the grain distribution used for each source as discussed in Sections 3.2.3 ($\mu$ Cep), 3.3 (VY CMa), and 3.4.2 (IRC +10420).}
\tablenotetext{b}{~Assumed average expansion velocity from widths of CO lines per \citet{De-Beck:2010}.}
\tablenotetext{c}{~Dynamical age of the shell $\Delta t = r_{2}$ / $v_{exp}$.  For IRC +10420 the average mass-loss rate is computed for two periods over the 6000 yr dynamical age of the shell, with the recent period spanning 2000 yr and the earlier period spanning 4000.   See Figure \ref{IRC_density_dist}.}
\tablenotetext{d}{~Average mass-loss rate computed using the adopted gas-to-dust mass ratio of 100:1 (see footnote \# 6 above).  The uncertainty is estimated from the range of optical depths that reproduce the FORCAST 30 $-$ 40 $\micron$ fluxes to within $\pm$20\%.}
\end{deluxetable*}
\end{turnpage}
\clearpage

\end{document}